\documentclass[11pt]{article}
\pdfoutput=1
\usepackage[utf8]{inputenc}
\usepackage{amsmath}
\usepackage{amsfonts}
\usepackage{float}
\usepackage{bm}
\usepackage{bbm}
\usepackage{natbib}
\usepackage{graphicx}
\graphicspath{ {./art/} }
\usepackage{subfig}
\usepackage{amsthm}
\usepackage{url}
\usepackage{xcolor}
\usepackage{setspace}
\usepackage[margin=1in]{geometry}
\usepackage{multirow}

\newtheorem{theorem}{Theorem}
\newtheorem{proposition}{Proposition}
\def\T{{ \mathrm{\scriptscriptstyle T} }}
\def\v{{\varepsilon}}

\title{Double Spike Dirichlet Priors for Structured Weighting}
\author{Huiming Lin and Meng Li}
\date{Department of Statistics, Rice University}

\begin{document}

\maketitle

\doublespacing

\begin{abstract}
Assigning weights to a large pool of objects is a fundamental task in a wide variety of applications. In this article, we introduce the concept of \emph{structured} high-dimensional probability simplexes, in which most components are zero or near zero and the remaining ones are close to each other. Such structure is well motivated by (i) high-dimensional weights that are common in modern applications, and (ii) ubiquitous examples in which equal weights---despite their simplicity---often achieve favorable or even state-of-the-art predictive performance. This particular structure, however, presents unique challenges partly because, unlike high-dimensional linear regression, the parameter space is a simplex and pattern switching between partial constancy and sparsity is unknown. To address these challenges, we propose a new class of double spike Dirichlet priors to shrink a probability simplex to one with the desired structure. When applied to ensemble learning, such priors lead to a Bayesian method for structured high-dimensional ensembles that is useful for forecast combination and improving random forests, while enabling uncertainty quantification. We design efficient Markov chain Monte Carlo algorithms for implementation. Posterior contraction rates are established to study large sample behaviors of the posterior distribution. We demonstrate the wide applicability and competitive performance of the proposed methods through simulations and two real data applications using the European Central Bank Survey of Professional Forecasters data set and a data set from the UC Irvine Machine Learning Repository (UCI). 
\end{abstract}


\section{Introduction}
There is an immense variety of applications in which the key is to assign weights to a large pool of objects, such as in model averaging, forecast combination, and tree ensembles. 
High-dimensional models and sparse learning methods have gained extensive attention in the literature; however, much less focus has been placed on the simplex setting, where the entries of the unknown parameter or weights are non-negative and sum up to one. 
In this paper, we are interested in \emph{structured} high-dimensional simplexes with \emph{sparsity} and \emph{partial constancy}, i.e., most of the weights are zero or near zero and the remaining ones are close to each other. 

Such structure is well motivated by the recurrent observation in various areas that equal weights often achieve favorable or even state-of-the-art empirical performance in practice, and the weights may be high-dimensional in modern applications. 

The first notable example is 
forecast combination in economics. There is a rich literature on how to optimally combine individual forecasts, and weights on the probability simplex are commonly used as they lead to unbiased estimation if each forecast is unbiased~\citep{bates1969combination,granger1984improved}. One remarkable observation is that the simplest combination method using equal weights constantly outperforms more complicated weighting schemes in terms of mean-squared forecast error, a phenomenon that is known as the `forecast combination puzzle'~\citep{clemen1989combining, stock2004combination, smith2009simple,Makridakis2020}. On the other hand, the number of forecasters can be large compared to the number of historical records in many applications, pointing to the high-dimensional regime where sparsity may be desired. 

Another example that may benefit from such structure is the well-known random forests method \citep{breiman2001random}, an ensemble method that first builds then averages a large collection of de-correlated trees. Random forests have long been considered as one of the most successful general-purpose supervised machine learning methods~\citep{svetnik2003random,segal2004machine,palmer2007random,biau2016random}. It is a special case of high-dimensional forecast combination with each tree regarded as an individual forecast.
\cite{breiman2001random} proved that the generalization error converges almost surely to a limit as the number of trees goes to infinity, which implied that above a certain number of trees, adding more trees in the forest does not improve accuracy much. This has invited discussions on whether a subset of individual trees can outperform the whole forest in prediction~\citep{kulkarni2012pruning}, in addition to its apparent advantage of improved parsimony. For example, \cite{zhang2009search} and \cite{bernard2009selection} developed methods to reduce the forest size while maintaining the prediction accuracy comparable to the initial forest. However, they did not incorporate both sparsity and partial constancy into ensemble methods such as random forests, which may lead to substantial predictive gain using real world data. Furthermore, the joint statistical inference on this constrained high-dimensional space with uncertainty quantification has not been studied in the literature. 

Such structure, high-dimensional simplex with partial constancy and sparsity, is indeed ubiquitous as equal weights and high-dimensional settings that necessitate sparsity are routinely encountered in modern applications, and convex combination often enjoys stability and theoretical guarantees~\citep{bunea2007aggregation,polley2010super}. Throughout this article, we assume the true parameter obeys this structure. While structured simplexes can be integrated into any model or flexible loss functions beyond the likelihood as long as the parameter space is a probability simplex, we will focus on high-dimensional ensemble learning for concreteness. This setting has wide applications; for example, it is a common framework in forecast combination~\citep{granger1984improved,diebold1990use}, and it serves as a main building block for \emph{super learner} in the model aggregation literature~\citep{van2007super,polley2010super}. 

When the high-dimensional parameters are constrained on a simplex, the popular $\ell_1$ penalty for sparse learning in high-dimensional regression falls short, as the $\ell_1$ norm of the parameters is equal to one. To circumvent this difficulty and achieve sparsity, one may alternatively penalize certain transformations of the parameters, say their square root.
However, solving such constrained optimization problem may be far from trivial as any regularizer that promotes exact sparsity under simplex constraints cannot be convex~\citep{li2020methods}.
There is recent literature on simplex structure, including sparse projection onto the simplex~\citep{kyrillidis2013sparse} and constrained optimization~\citep{clarkson2010coresets}. In the high-dimensional setting, \citet{li2020methods} discussed strategies to estimate sparse simplexes, including to couple empirical risk minimization with thresholding or re-weighted $\ell_1$ regularization. 
\citet{conflitti2015optimal} studied the high-dimensional forecast combination problem and found that the probability simplex constraint helped stabilize the determination of the optimal weights. However, the constraint of partial constancy is not studied, and uncertainty quantification remains a challenge. 

A natural approach to carry out estimation constrained on the structured simplex space is to adopt a two-step strategy of `sparsity first, then partial constancy': first select a subset of forecasts, then shrink the remaining ones toward a constant while maintaining the simplex constraint. There is a rich menu of methods for the first step, for example, lasso~\citep{tibshirani1996regression}; this approach has been implemented in \citet{diebold2019machine}, and hereafter we refer to it as two-step lasso. However,
such a two-step strategy suffers from the model selection error in the first step, such as false positives of lasso~\citep{bogdan2015slope,su2017false}, which is disruptive for estimating structured simplexes in light of the strong structure implied by partial constancy. Moreover, there is no guarantee that the selected forecasters always have non-negative coefficients, leading to ambiguity regarding whether coefficients of opposite signs should be reset to the same constant. Indeed, our simulations show that two-step lasso may yield a non-negligible estimation error even in the most ideal case when the parameter strictly follows the structure, and its performance is highly sensitive to its regularization parameter. 

Using Dirichlet priors in the Bayesian paradigm enables a unified approach to learn simplexes with uncertainty quantification. 
Along the line of weighting objects, \cite{yang2014minimax} considered convex aggregation and assigned symmetric Dirichlet priors to aggregation weights, and \cite{rousseau2011asymptotic} studied overfitted mixture models where they advocated using Dirichlet distributions with small concentration parameters as priors for weights of densities.
There is also a rich literature on generalizing Dirichlet distributions via mixtures to incorporate richer dependence patterns; see, for example, \cite{ongaro2013generalization, migliorati2017structured,ongaro2020new}. In a recent work, \cite{heiner2019structured} proposed a sparsified Dirichlet mixture prior to retain sparsity in probability simplex vectors. None of the Dirichlet priors in these works account for partial constancy, a key characteristic of our weight parameter space that arises from the aforementioned motivation.

In this article, we propose a Bayesian method to simultaneously accommodate sparsity, partial constancy, and the simplex constraint under the high-dimensional setting, while allowing for inference. This extends the useful structure embedded in the `forecast combination puzzle' to a broader class of applications. At the core of our method lies a new class of \emph{double spike Dirichlet} priors, which leads to joint inference without resorting to two-step alternatives that have pitfalls.

When applied to ensemble learning, the proposed double spike Dirichlet priors lead to a principled Bayesian strategy for structured ensembles, which enables uncertainty quantification and tuning key parameters without relying on cross-validation. We establish a posterior contraction property by showing the entire posterior distribution concentrates around the desired structure, adaptive to the unknown sparsity level, which to the best of our knowledge is the first of its kind for structured high-dimensional probability simplexes. The proposed method is computationally easy to implement using Metropolis-Hastings algorithms. We illustrate through simulations and two case studies that the proposed method leads to substantial performance gain over alternatives. In particular, the proposed method complements random forecasts with improved prediction, enhanced parsimony, and easy implementation, strongly suggesting its wide applicability building on the success and popularity of random forests in practice. We provide R code at \url{https://github.com/xylimeng/StructuredEnsemble} for routine implementation. 

The rest of the paper is organized as follows. Section~\ref{sec:DSD} presents the proposed double spike Dirichlet priors and compares their properties to that of the symmetric Dirichlet prior. The posterior sampling algorithm is described in Section~\ref{sec:PS}. Section~\ref{sec:PP} studies properties of the coefficients' posterior distribution and gives a posterior contraction rate. Simulation studies are carried out in Section~\ref{sec:simu}. In Section~\ref{Sec:realdata} and Section~\ref{sec:RF}, the method is illustrated by real data applications using the European Central Bank Survey of Professional Forecasters data set and the concrete data set ~\citep{yeh1998modeling} from the UCI repository~\citep{Dua:2019}, respectively.  

\section{The Double Spike Dirichlet Priors}
\label{sec:DSD} 
We consider a high-dimensional ensemble learning problem with model weights constrained on a simplex.
Let $\Theta^{K-1} = \{ \beta\mid \beta\in \mathbb{R}^K,\sum_{i =1}^K \beta_i = 1, \beta_i \geq 0, 1 \leq i \leq K\}$
be a probability simplex. Ensemble learning can be cast as a constrained linear regression model
\begin{equation*}
    Y=X\beta+\v,
\end{equation*}
where the unknown $K$-dimensional weight parameter $\beta$ is constrained on the probability simplex $\Theta^{K-1}$, $X$ is the given $n \times K$ design matrix with each column being the prediction by a forecaster or learner, and $\v \sim N(0,\sigma^2I)$ is the error term with $\sigma^2$ possibly unknown. We allow the situation $n \leq K$ and are interested in structured coefficients, i.e., 
we assume the ground truth $\beta^*$ belongs to the set 
\begin{equation} \label{eq:structure}
    \begin{split}
        \Theta(s,K)=\{\beta\;\vert\;\beta \in \Theta^{K-1} &\text{  has only $s\ge 1$ nonzero elements} \\ 
        &\text{  and they all take a common value $1/s$}\},
    \end{split}
\end{equation}
where the number and locations of nonzero elements are both unknown, and $s$ is considerably smaller than $K$. We note that the methods developed in this article can generalize to any model or risk functions in lieu of the Gaussian likelihood, such as generalized linear regression and Gibbs posterior~\citep{jiang2008gibbs}, leading to structured counterparts.

Motivated by the special structure  in Equation~\eqref{eq:structure} as well as the simplex constraint, we propose the following hierarchical class of double spike Dirichlet priors on $\beta$:
\begin{equation}\label{eq:DSD}
\begin{split}
    &\beta |\gamma,\rho_1,\rho_2 \sim  \text{Dir}\left(\rho_1\gamma+\rho_2(1-\gamma)\right)\\
    &\gamma_i |\theta \overset{\text{independent}}\sim \text{Bernoulli}(\theta),\quad i=1,\ldots, K,
\end{split}
\end{equation}
where $\rho_1 >\rho_2>0$. We use a conjugate inverse-Gamma prior on $\sigma^2$ when it is unknown, i.e., $\sigma^{-2}|a_1,a_2 \sim \text{Gamma}(a_1,a_2)$.  
In Equation~\eqref{eq:DSD}, $\gamma_i=1$ indicates $\beta_i$ is associated with the larger concentration parameter $\rho_1$, and otherwise a smaller concentration parameter $\rho_2$. Let $\gamma=(\gamma_1,\ldots,\gamma_K)$ and denote $|\gamma| = \sum_{i=1}^{K}\gamma_i$. Marginalizing out $\gamma$, the prior on $\beta$ becomes
$$
g(\beta;\theta,\rho_1,\rho_2)=\sum_{\gamma}\frac{\Gamma\left(\rho_1|\gamma|+\rho_2(K-|\gamma|)\right)}{\Gamma(\rho_1)^{|\gamma|}\Gamma(\rho_2)^{K-|\gamma|}}\prod_{i=1}^{K}\beta_i^{\rho_1\gamma_i+\rho_2(1-\gamma_i)-1}\theta^{|\gamma|}(1-\theta)^{K-|\gamma|}.
$$
The proposed prior is a mixture of Dirichlet distributions, where the concentration parameters are independent Bernoullis supported on $\{\rho_1, \rho_2\}$.  
Through this article, we shall choose a large value for $\rho_1 \gg 1$ and a small value for $\rho_2 \ll 1$ to encode the structural information formulated in Equation~\eqref{eq:structure}, although the posterior sampling method in Section~\ref{sec:PS} is applicable for any $(\rho_1, \rho_2)$. With such choices, conditional on $\gamma$, elementary properties of Dirichlet distributions indicate that those $\beta_i$ associated with $\gamma_i=0$ tend to be near 0 and the remaining ones are nonzero and tend to be near $1/|\gamma|$, leading to two spikes around which the values of $\beta$ concentrate. Hence, we call $g(\beta;\theta,\rho_1,\rho_2)$ the double spike Dirichlet priors. In the hierarchical representation of $g(\beta;\theta,\rho_1,\rho_2)$ in \eqref{eq:DSD}, the use of Dirichlet distributions enables inference on parameters constrained to $\Theta(s, K)$ in a unified manner. In contrast, two-step alternatives discussed in the preceding section can be loosely interpreted as learning $\gamma$ via variable selection methods, followed by post-hoc normalization.

Employing the well-known relationship between Dirichlet and Gamma distributions, an equivalent form of the double spike Dirichlet priors is
\begin{equation}\label{eq:rescaled gamma form}
\begin{split}
    \beta&=\left(\sum_{i=1}^p u_i\right)^{-1}u,\quad u=(u_1,\ldots,u_p), \quad u_i|\gamma_i\overset{\text{independent}}\sim\text{Gamma}(\gamma_i\rho_1+(1-\gamma_i)\rho_2, 1),\\
    \gamma_i&|\theta \overset{\text{independent}}\sim \text{Bernoulli}(\theta), \quad i=1,\ldots,K.
\end{split}
\end{equation}
Hence, the double spike Dirichlet prior dichotomizes the shape parameter in the Gamma distributions. This representation facilitates posterior sampling as shown in the next section.

It is well known that the concentration parameters in the Dirichlet distribution are instrumental in determining prior properties, the choice of which has invited an extensive discussion. When using Dirichlet priors to learn multinomial distributions, for example, \cite{good1967bayesian} proposed a hierarchical approach by assigning priors to these parameters, and \cite{fienberg1973simultaneous} discussed a variety of choices for the sum of concentration parameters. For weighting objects using Dirichlet priors, the symmetric Dirichlet distribution $(\pi_1,\ldots, \pi_K) \sim \text{Dir}(\alpha,\ldots, \alpha)$, which can be seen as a special case of our prior with $\rho_1=\rho_2$, is still a default prior on the probability simplex. Often a very small $\alpha$ value is chosen to induce sparsity. However, as indicated by the following proposition, as $\alpha \to 0$, most realizations sampled from the symmetric $\text{Dir}(\alpha,\ldots, \alpha)$ are very close to vectors containing all zeros with a single one at a random location from $1,\ldots, K$.
\begin{proposition}\label{lemma}
Let $(\pi_{(1)},\ldots, \pi_{(K)})$ be the order statistics of $(\pi_1,\ldots, \pi_K)$. If $(\pi_1,\ldots, \pi_K) \sim \text{Dir}(\alpha,\ldots, \alpha)$, then for any $t \in (0,1)$ and $\alpha>0$, we have 
$$
\text{pr}(\pi_{(K-1)}\le t\pi_{(K)})\ge t^{\alpha(K-1)}.
$$
\end{proposition}
\begin{proof}
See the Appendix.
\end{proof}
Proposition \ref{lemma} indicates that for any $t\in (0,1)$ and $\alpha>0$, we have 
$$
\text{pr}(\pi_{(K-1)}\le t)\ge \text{pr}(\pi_{(K-1)}\le t\pi_{(K)})\ge t^{\alpha(K-1)}.
$$
Therefore, if we let $\alpha \downarrow 0$, the probability $\text{pr}(\pi_{(K-1)}\le t) \to 1$, which implies that the second
largest probability $\pi_{(K-1)}$ is bounded from above by any fixed constant $t$ with probability approaching 1. Furthermore, for small $\alpha \in (0,1)$, if we let $t=\alpha$, we obtain that
$$
\text{pr}(\pi_{(K-1)}\le \alpha)\ge \alpha^{\alpha(K-1)}.
$$
If $\alpha=o(1)$, it follows that
$$
\text{pr}(\pi_{(K-1)}\ge \alpha)\le 1-\alpha^{\alpha(K-1)}=1-e^{\alpha(K-1)\log \alpha}=O(-\alpha\log \alpha)=o(\alpha^c),
$$
for any positive constant $c<1$. 
In contrast, the proposed double spike Dirichlet prior
puts approximately at least $\binom{K}{s} \theta^s(1-\theta)^{K-s}$ mass on probability vectors with any number $s$ of nonzeros for small $\rho_2$ and large $\rho_1$. This is formally described by the following proposition. 
\begin{proposition}\label{lem2}
For any $s \in \{1, \ldots, K\}$, let $N$ be the number of $\beta_i$ that are greater than $1/(2s)$ under the double spike Dirichlet prior. Suppose $\rho_2 \leq 1/K$ and $\rho_1\ge 5/s$. Then we have 
\begin{equation} \label{eq:prior.tail}
\text{pr}(N\ge s)\ge \binom{K}{s}
\theta^s(1-\theta)^{K-s}(1-9s/\rho_1).
\end{equation}
\end{proposition}

\begin{proof}
See the Appendix.
\end{proof}

The tail probability in Equation~\eqref{eq:prior.tail} is close to $\binom{K}{s} \theta^s(1-\theta)^{K-s}$ for sufficiently large $\rho_1$, which is the probability of the event $\{|\gamma| = s\}$. 
The configuration of samples from the proposed prior depends on the choice of $\theta$, $\rho_1$ and $\rho_2$, where a large $\rho_1$ together with a small $\rho_2$ helps shrink the nonzeros to a constant. With $\rho_1\to \infty$ and $\rho_2\to0$, the proposed prior approaches  $\beta_i=\gamma_i/\sum_{i=1}^K\gamma_i$ conditional on $\gamma$, corresponding to the exact structure in $\Theta(s, K)$. For other choices of $(\rho_1, \rho_2)$, the proposed prior is a continuous relaxation. This flexibility is motivated by the consideration that many weights might be small (but not exactly zero) while the others are close to each other (but not exactly equal to each other), similar to the rich but different literature on continuous shrinkage priors in sparse high-dimensional regression. The developed sampling algorithm and theory in this article are applicable for both the limiting case and continuous relaxation.

\section{Posterior Sampling and Posterior Summary}
\label{sec:PS}
We adapt the stochastic search algorithm that used add, delete and swap moves for posterior sampling~\citep{brown1998bayesian}. We add a stay step so that the algorithm could keep drawing samples of $\beta$ to improve mixing if stabilizing around a vector of $\gamma_i$'s. As such, we propose a random search algorithm for posterior sampling, which we call Add/Delete/Swap/Stay (ADSS). In the literature, methods based on this random walk type of Markov chain Monte Carlo (MCMC) are commonly used for posterior sampling of multivariate or high dimensional binary vectors~\citep{chapple2017bayesian}. The algorithm is essentially Metropolis-Hastings; it exploits the well-known relationship between the Dirichlet and Gamma distribution as described in \eqref{eq:rescaled gamma form}.

The ADSS algorithm includes the following steps:
\begin{enumerate}
    \item Set $t$=0. Initialize $(\sigma^{-2})^{(t)}$, $\gamma_i^{(t)}$, $A_i^{(t)}$ for $i=1,\ldots, K$. Set $\beta^{(t)}=A^{(t)}/\left\Vert A^{(t)}\right\Vert_1$, where $A^{(t)}=(A_1^{(t)},\ldots, A_K^{(t)})$, and $\left\Vert\cdot\right\Vert_1$ is the $\ell_1$ norm of a vector.
    \item Set $t=t+1$. Given $\gamma^{(t-1)}$, initialize a candidate vector $\tilde{\gamma}=\gamma^{(t-1)}$. Proceed to one of the following with equal probability:
    \begin{enumerate}
        \item (add) randomly select a $j$ from $J=\{j'\vert \gamma_{j'}^{(t-1)}=0\}$. Set $\tilde{\gamma}_j=1$;
        \item (delete) randomly select a $j$ from $J^c=\{j'\vert \gamma_{j'}^{(t-1)}=1\}$. Set $\tilde{\gamma}_j=0$;
        \item (swap) randomly select a $j_1$ from $J$ and $j_2$ from $J^c$.
    Set $\tilde{\gamma}_{j_1}=1$, $\tilde{\gamma}_{j_2}=0$;
        \item (stay) no actions;
    \end{enumerate}
    Conditional on $\tilde{\gamma}$, for $i=1,\cdots, K$, propose a candidate $\tilde{A}_i \sim \text{Gamma}(\rho_1\tilde{\gamma}_i+\rho_2(1-\tilde{\gamma}_i),1)$. Set $\tilde{\beta}=\tilde{A}/\left\Vert \tilde{A}\right\Vert_1$, where $\tilde{A}=(\tilde{A}_1,\cdots,\tilde{A}_K)$.
    
    \item  Accept $\gamma^{(t)}=\tilde{\gamma}$ and  $\beta^{(t)}=\tilde{\beta}$
    with probability
    $$
    \min{\left(1,\frac{\{\theta/(1-\theta)\}^{|\tilde{\gamma}|}}{\{\theta/(1-\theta)\}^{|\gamma^{(t-1)}|}}\frac{\exp\{-(\sigma^{-2})^{(t-1)}\sum_{i=1}^n(y_i-x_i^{T}\tilde{\beta})^2/2\}}{\exp\{-(\sigma^{-2})^{(t-1)}\sum_{i=1}^n\left(y_i-x_i^{{T}}\beta^{(t-1)}\right)^2/2\}} \right)};
    $$
    
    Otherwise, keep $\gamma^{(t)}=\gamma^{(t-1)}$, $\beta^{(t)}=\beta^{(t-1)}$.
    
    \item Draw $(\sigma^{-2})^{(t)} \sim \text{Gamma}\left(a_1+n/2, a_2+\sum_{i=1}^n\left(y_i-x_i^{{T}}\beta^{(t)}\right)^2/2\right)$.
    
    \item Repeat Steps 2--4 for $niter$ times.
\end{enumerate}

Remark. For the limiting case when $\rho_1\to\infty$ and $\rho_2\to 0$, which encodes the exact structure $\beta_i=\gamma_i/|\gamma|$ in the prior, the ADSS algorithm proceeds by setting $A_i^{(t)}=\gamma_i^{(t)}$ in Step 1 and $\tilde{A}_i=\tilde{\gamma}_i$ in Step 2.

A Gibbs sampler is an alternative sampling strategy to ADSS, but it faces increasing challenges partly due to the simplex constraint. In a conventional Gibbs sampler, to update a single $\gamma_i$, one may draw a sample from the Bernoulli distribution with success probability $\frac{\theta A_i^{\rho_1}/\Gamma(\rho_1)}{\theta A_i^{\rho_1}/\Gamma(\rho_1)+(1-\theta) A_i^{\rho_2}/\Gamma(\rho_2)}$, which is usually close to 0 for all $i$. To improve mixing, a more appealing implementation of the Gibbs sampler would integrate out these $A_i$ and draw directly from $\gamma_i|Y$. However, this is challenging when $\beta$ is constrained on the simplex, unlike in the usual linear regression models where $\beta$ is unconstrained and the integration can be analytically tractable. Therefore, we do not pursue Gibbs sampling and instead use a proposal distribution and Metropolis-Hastings for posterior sampling.

The ADSS algorithm is amenable to a fully Bayesian approach for $\theta$ by giving it a prior and sampling it from the posterior distribution. In particular, by specifying a $\text{Beta}(b_1, b_2)$ prior for $\theta$ and exploiting conjugacy, the Gibbs update formula for $\theta$ at step $t$ is to draw $\theta^{(t)}\sim \text{Beta}(b_1+|\gamma^{(t)}|, b_2+K-|\gamma^{(t)}|)$. For hyperparameters, one may choose $b_1=1$ and $b_2=0.5K$ as suggested by \cite{rovckova2018spike}.

The posterior summary depends on the inferential goal. For example, if a point estimate of $\beta$ is of interest, we may report the posterior mean after the burn-in period. If the uncertainty of the estimated $\beta$ is also of interest, we may report a region of $\beta$ based on quantiles of $\ell_1$ errors. For the task of prediction such as in our real data applications, we mainly use the posterior mean of $\beta$ to combine forecasts, unless stated otherwise. 

\section{Posterior Properties}
\label{sec:PP}
In this section, we study posterior properties of $\beta$, which guarantees the entire posterior distribution concentrates around the true structure and leads to insights about choosing $\rho_1, \rho_2$ and $\theta$. For notational simplicity, we assume $\sigma=1$. We first introduce a compatibility condition on the design matrix. Define the compatibility number of the model by
$$
\Phi(s)=\inf\left\{\frac{\left\Vert X\beta\right\Vert_2\surd{s}}{\left\Vert X\right\Vert\left\Vert\beta\right\Vert_1}:\left\Vert\beta\right\Vert_1\leq 2\right\},
$$
where $\left\Vert X\right\Vert=\max_{j}\left\Vert X_{\cdot,j}\right\Vert_2$, and $\left\Vert \beta\right\Vert_q=(\sum_{j}|\beta_j|^q)^{1/q}$ for $1\leq q\le \infty$.
We say the model satisfies the compatibility condition if $\Phi(s)>0$. The compatibility condition ensures that the parameter $\beta$ is estimable; in fact, the model may not be identifiable when $K>n$ if we do not add constraints to the design matrix. Similar conditions are used in the high-dimensional regression literature when $\beta$ is not constrained on the simplex; for example, see~\cite{buhlmann2011statistics, castillo2015bayesian}.

The following theorem asserts that under certain conditions, the posterior distribution concentrates around the ground truth $\beta^*$ at a certain rate.

\begin{theorem}\label{maintheorem}
Suppose $0<s=o(K)$, $\rho_1=K^{\alpha_1}$, $\rho_2=K^{-\alpha_2}$, $\alpha_1, \alpha_2 >0$, $\alpha_1/2+\alpha_2\geq 1$. Under compatibility condition $\Phi(s)>0$, if $\theta$ is chosen such that $O(1)/K\le\theta\leq s/K$, then for sufficiently large $M>0$, we have
\begin{equation*}
    \sup_{\beta^* \in \Theta(s,K)}E_{\beta^*}\Pi\left(\left\Vert\beta-\beta^*\right\Vert_1 > \frac{M}{\Phi(s)}\frac{s\log K}{\min(\left\Vert X\right\Vert,\surd{K^{\alpha_1}})}|Y\right)\to 0.
\end{equation*}
\end{theorem}

\begin{proof}
See the Appendix.
\end{proof}

Remark 1. \citet{castillo2015bayesian} studied full Bayesian procedures for high-dimensional linear regression under sparsity constraints.  Under their assumptions, they showed that $$ \sup_{\beta^*}E_{\beta^*}\Pi\left(\left\Vert\beta-\beta^*\right\Vert_1 > \frac{M}{\bar{\psi}(s)^2}\frac{s\surd{\log K}}{\left\Vert X\right\Vert\phi(s)^2}|Y\right)\to 0,$$ 
where $\psi(s)$ and $\phi(s)$ are related to their compatibility conditions. Although our rate has a similar form, the proof is radically different due to the structured simplex constraint under consideration.

Remark 2. Theorem~\ref{maintheorem} reveals that both $\left\Vert X\right\Vert$ and $K^{\alpha_1}$ affect the posterior contraction rate, which depicts the effect of the likelihood and prior, respectively. Indeed, it can be deduced from the proof that a larger $\left\Vert X\right\Vert$ is linked to a stronger likelihood, while a larger $K^{\alpha_1}$ leads to a tighter thus stronger prior. In particular, when the likelihood is strong enough to identify the nonzero coefficients, a stronger prior helps enforce the nonzero coefficients to be equal. The parameter $\alpha_2$ does not affect the contraction rate in Theorem~\ref{maintheorem} as long as $\alpha_2\geq 1-\alpha_1/2$. 

Remark 3. To achieve the rate in the theorem, the choice of $\theta$ can be very flexible: instead of requiring $\theta = s/K$, letting $\theta = t/K$ for any $1\leq t \leq s$ leads to the established rate. That is, the proposed method is \emph{adaptive} to the unknown sparsity level.

\section{Simulations}\label{sec:simu}
We carry out simulations to investigate the finite-sample performance of the proposed Bayesian method. Section~\ref{sec:two.scenarios} focuses on two scenarios, one following the exact partially constant structure and the other with deviation, and assesses estimation accuracy and sensitivity to hyperparameters for both the proposed method and competing methods. Using extended comparisons, Section~\ref{sec:extended.sim} studies the effect of varying levels of deviation from the partially constant structure, different variants of the methods in comparison, and the effect of signal-to-noise ratio.

To implement our method, we use the ADSS algorithm to draw posterior samples. We use the following default settings in our numerical
experiments, unless otherwise stated. We choose $\rho_1=K^2$ and $\rho_2=1/K$, following the theoretical results in Section~\ref{sec:PP} and particularly the condition $\alpha_1/2+\alpha_2\ge 1$. We put a weak $\text{Gamma}(a_1,a_2)$ prior on $\sigma^{-2}$ with $a_1=a_2=0.01$. For posterior sampling for the proposed method and other Bayesian methods, we run $niter=20000$ iterations and burn in the first 15000. For the choice of $\theta$, we set it to $1/K$, which is supported by Theorem~\ref{maintheorem}, but we note that one can also specify other preferred values or a range of values based on their domain knowledge about the sparsity level. Other than the default choice of fixing $\theta$, we will also compare a fully Bayesian treatment by placing a Beta prior on $\theta \sim \text{Beta}(1, 0.5K)$. In our numerical experiments, we find that the ADSS algorithm is numerically stable without triggering rounding or overflow errors.

\subsection{Two Scenarios} \label{sec:two.scenarios} 
In the first scenario, the true simplex parameter strictly follows the sparse and partially constant structure, while the second scenario allows some deviation.

We consider two alternative methods for comparison: a Bayesian method using the symmetric Dirichlet prior and the two-step lasso method proposed by~\citet{diebold2019machine}. For the symmetric Dirichlet method, we use the algorithm in~\citet{yang2014minimax}. Tuning parameters in these two methods include the concentration parameter $\rho$ for symmetric Dirichlet, and the regularization parameter $\lambda$ for two-step lasso. We give them an advantage throughout the numerical experiments by searching a grid of hyperparameters and reporting the best result, as opposed to fixing the hyperparameter values when implementing our method. Details of the grid search are deferred to the sensitivity analysis later in this section. 

We use the $\ell_1$ error of the estimated $\beta$ as the metric to evaluate the three methods. For the two Bayesian methods, we calculate the $\ell_1$ error for each of the 5000 MCMC samples, and report the averaged $\ell_1$ error. This metric directly points to posterior contraction that is unique to Bayesian methods and has been reflected in the rate calculation in Theorem~\ref{maintheorem}, and utilizes the entire posterior samples without relying on point estimates, eliminating the dependence on posterior summary for which multiple strategies exist. For two-step lasso, we use its point estimate to calculate the $\ell_1$ error as uncertainty quantification for this non-Bayesian method is not immediately available. According to Jensen's inequality and the convexity of the $\ell_1$ norm, the reported $\ell_1$ error for the two Bayesian methods would be no smaller than the one calculated using the posterior mean; hence, our comparison gives the two-step alternative another advantage as the $\ell_1$ error of the two Bayesian methods could be further improved if a point Bayes estimator is used instead. Each scenario is repeated 100 times, and the averaged $\ell_1$ error is compared. 

For the first scenario, we generate 80 samples from $y=x^T\beta^*+\v$, where $x\in \mathbb{R}^{40}$, $\v \sim N(0,1.5^2)$, and $\beta^*=(1/3, 1/3, 1/3,0,\ldots, 0)$, i.e., $n=80$, $K=40$, $s=3$ in this case. Each column of the design matrix $X$ is generated independently from $N(0, 3^2)$. 

The proposed method yields an average $\ell_1$ error 0.032. The best $\ell_1$ error over all hyperparameter sets is 0.428 and 0.139 for the symmetric Dirichlet prior and two-step lasso, respectively. This suggests that under this simulation setting the proposed estimates are more accurate than alternatives by a substantial margin; recall that the alternatives have been equipped with the optimal hyperparameter value that achieves the smallest $\ell_1$ error. The performance gain of our method over symmetric Dirichlet is not surprising as the ground truth does follow the sparse and partially constant pattern as encoded in the double spike Dirichlet prior.

The second scenario mimics the first scenario but allows derivation of $\beta^*$ from the sparse and partially constant pattern. In particular, the last $K-s$ entries of $\beta^*$ equally share a total of $0.05$ weight, and the first $s$ entries are $(\beta_1, \beta_2, \beta_3)=(0.3089, 0.3672, 0.2739)$. Its closest projection to the partially constant simplex parameter space is the $\beta^*$ used in the first scenario, and their $\ell_1$ distance is 0.1677.

The proposed method yields an average $\ell_1$ error 0.210. The best $\ell_1$ error achieved is 0.436 for the symmetric Dirichlet prior and 0.298 for two-step lasso. Not surprisingly, both our method and two-step lasso have a larger $\ell_1$ error than the first scenario due to the deviation of $\beta^*$ from the exact structure. However, the proposed method appears to be more flexible and can better adapt to such cases than two-step lasso. On the other hand, the symmetric Dirichlet method continues to show a larger error than our method, perhaps due to the lack of a strong structural prior when the sample size is relatively small compared to the number of predictor variables.

We now turn to the mixing of $\gamma$ in the proposed method. Sampling of high-dimensional discrete indicators could be challenging if the target distribution is highly multimodal \citep{schafer2012monte}. Interestingly, the strong structural information that our double spike prior encodes appears to greatly mitigate such concerns. When the underlying $\beta^*$ follows (or approximately follows) the sparse and partially constant structure, the proposed prior tends to yield a posterior distribution that is less likely to be multi-modal. This is because unlike high-dimensional unconstrained parameter space, a small change in $\gamma$ could mean a large difference in $\beta$ and subsequently in the likelihood, when $\beta$ is constrained to the sparse, partially constant simplex. For example, the $\ell_1$ distance between any two elements in $\Theta(s, K)$ as well as that between $\Theta(s, K)$ and $\Theta(s - 1, K)$ is lower bounded by $2/s$. To see this, we construct elements such that the lower bound is achieved. In particular, we consider $(1/s, \ldots, 1/s, 0, \ldots, 0) \in \Theta(s, K)$ versus any other element in $\Theta(s, K)$ with one of its last $K - s$ element being $1/s$, and $(1/s, \ldots, 1/s, 0, \ldots, 0) \in \Theta(s, K)$ versus $(1/(s - 1), \ldots, 1/(s - 1), 0, \ldots, 0) \in \Theta(s - 1, K)$. However, these distances could be arbitrarily small if the parameter $\beta$ is unconstrained as in traditional high-dimensional linear regression settings. 

To demonstrate the aforementioned tendency toward uni-modality using our numerical experiments, we take the first scenario as an example and assess the relative frequency of two different $\gamma$'s by computing an approximated posterior probability ratio $p(\gamma|Y)/p(\tilde{\gamma}|Y)$ for $\gamma\neq\tilde{\gamma}$. It is equivalent to use the unnormalized $p(\gamma|Y)$, which, after integrating out $\sigma^2$, is approximately $p(\gamma|Y)\propto \{a_2+\lVert Y-X\frac{\gamma}{|\gamma|}\rVert_2^2/2\}^{-(a_1+n/2)}\theta^{|\gamma|}(1-\theta)^{K-|\gamma|}.$
This approximation is obtained by considering the limiting case when $\rho_1\to\infty$ and $\rho_2\to0$.
We look into one generated data set and find that there are 10 unique $\gamma$ vectors in the posterior samples; after burn-in, there is only one unique $\gamma$, which is $\gamma^*=(1,1,1,0,0,\ldots,0)$. For any $\gamma\neq \gamma^*$ in the 10 unique vectors, the highest posterior probability ratio ${p(\gamma|Y)}/{p(\gamma^*|Y)}=3.49 \times 10^{-7}$, which is very small. This provides strong evidence against multi-modality that hurdles mixing of an MCMC sampler. In Section~\ref{sec:RF}, we further assess the convergence of our MCMC samples using real data when the exact structure is not guaranteed to hold. 

We next provide sensitivity analyses for the three methods with respect to hyperparameters. For the proposed method, we compare results for $\rho_1$ = \{6.3, 23.6, 88.2, 329.2, 1229.4, 4590.4, 17140.2, 64000\} and $\rho_2=\{1.6 \times 10^{-5}, 6.8 \times 10^{-5}, 3 \times10^{-4}, 1.3\times10^{-3}, 5.7\times10^{-3}, 2.5\times10^{-2}\}$, which are generated by taking $K$ to the power of an equally-spaced grid on $[0.5,3]$ and $[-3, -1]$, respectively. For the symmetric Dirichlet with the concentration parameter $\rho$, we vary $\rho$ from the set for $\rho_2$. For the regularization parameter $\lambda$ in two-step lasso, we use a grid of 80 values equally spaced on $[-8, 8]$ for $\log \lambda$, and this gives a list of $\lambda$ ranging from 0.0003 to 2981.

\begin{figure}[tbp]
\minipage{0.48\textwidth}
\includegraphics[width = \linewidth]{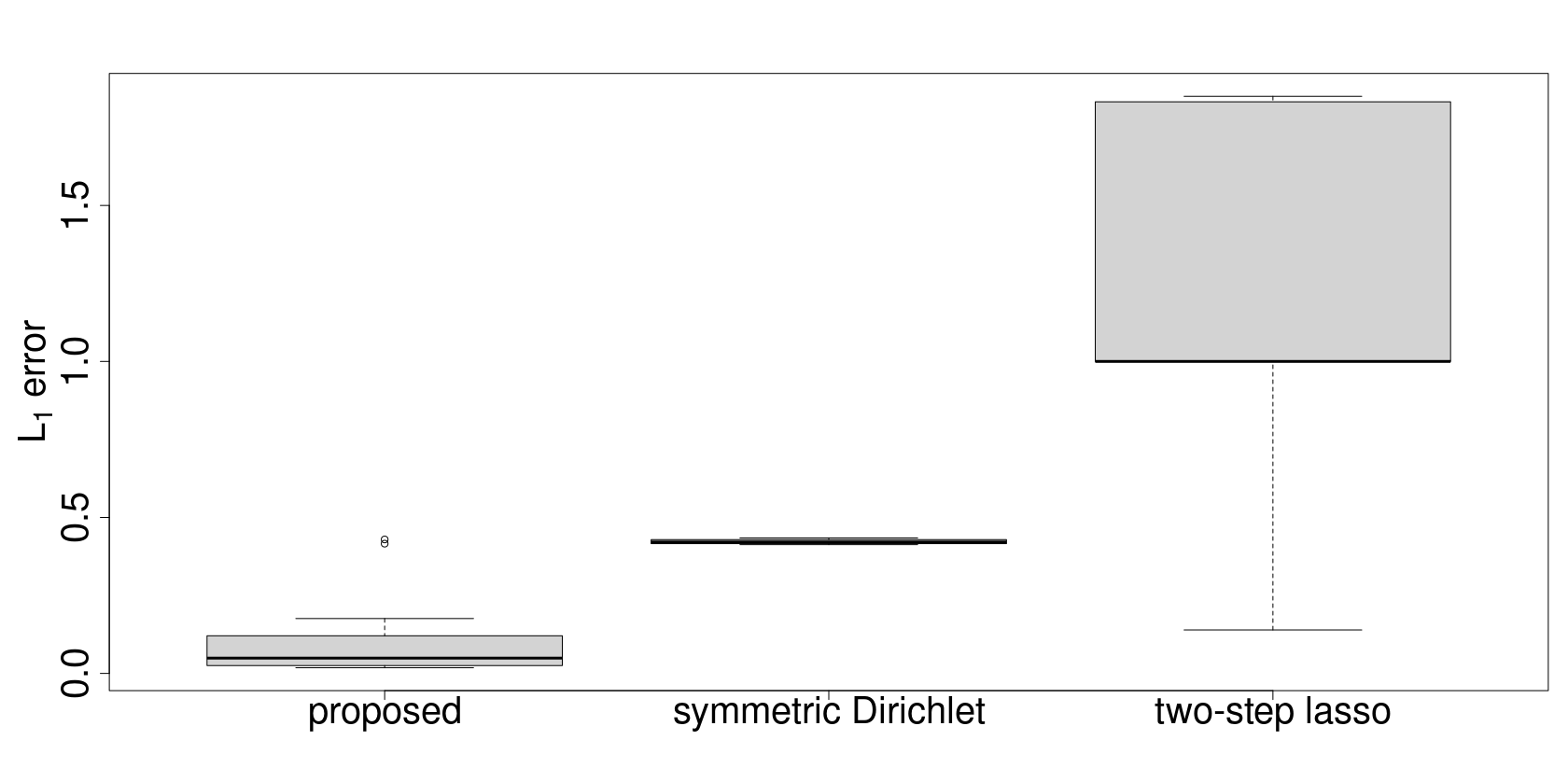}
\endminipage\hfill
\minipage{0.48\textwidth}
\includegraphics[width=\linewidth]{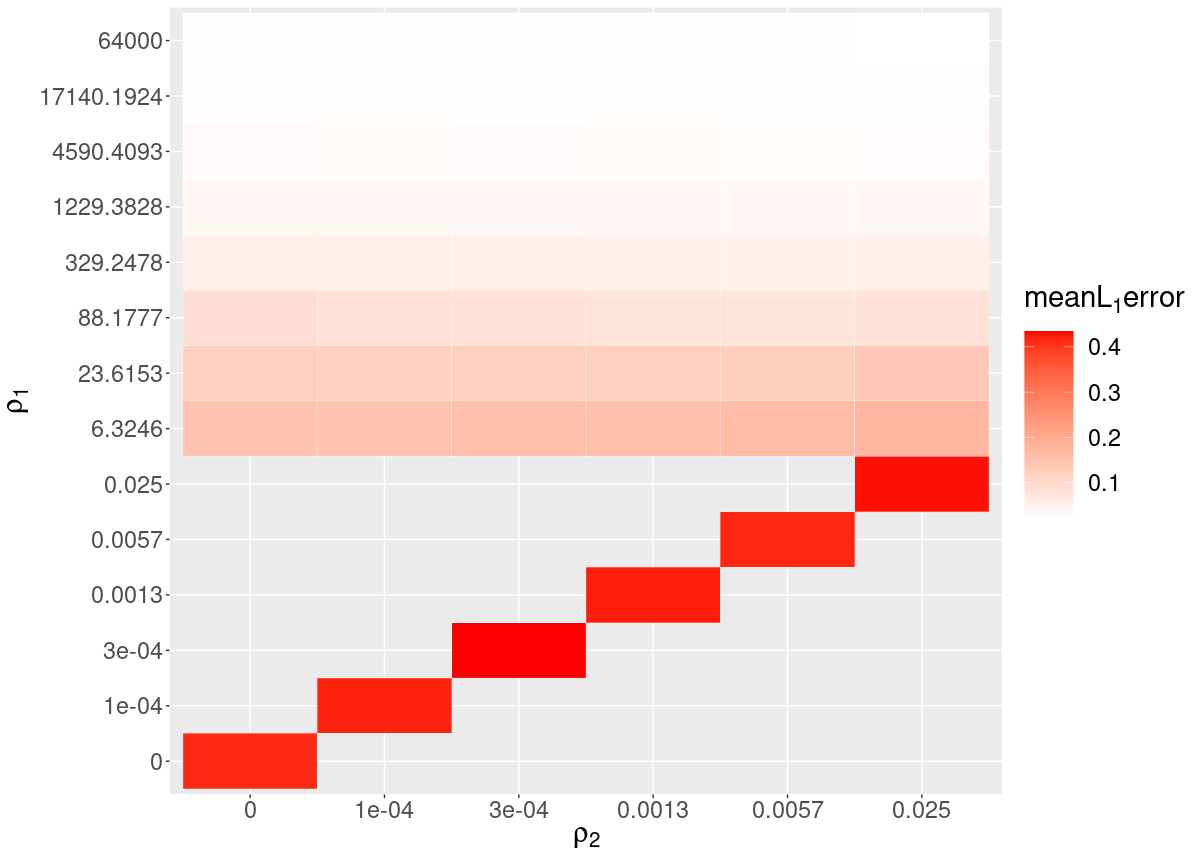}
\endminipage\hfill
\minipage{0.48\textwidth}
\includegraphics[width=\linewidth]{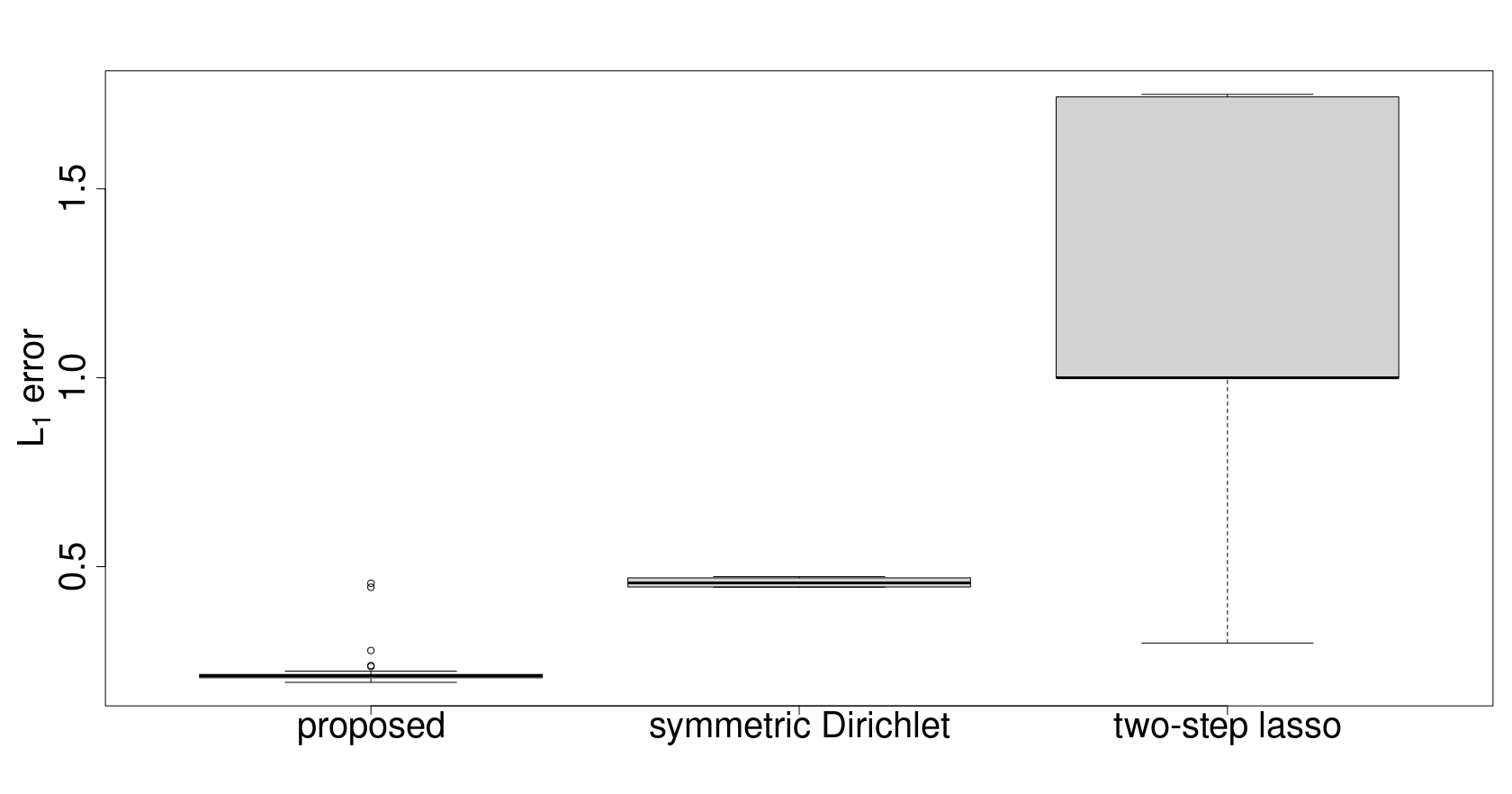}
\endminipage\hfill
\minipage{0.48\textwidth}
\includegraphics[width=\linewidth]{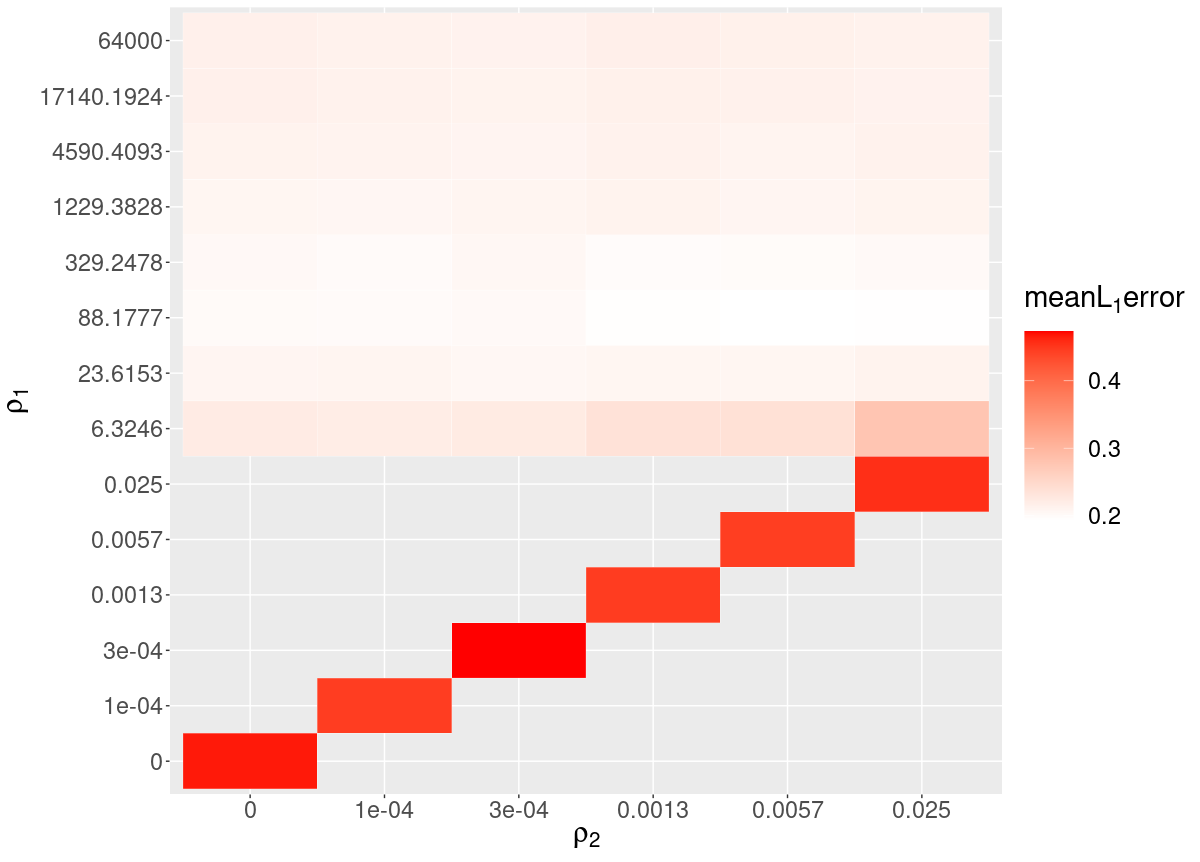}
\endminipage
\caption{Comparison of the $\ell_1$ error of coefficient estimates of three methods averaged over 100 simulations for Scenario 1 (the first row) and Scenario 2 (the second row). Left: boxplots of $\ell_1$ error as hyperparameters vary (the hyperparameters are $(\rho_1, \rho_2)$ for the proposed method, $\rho$ for the symmetric Dirichlet prior, and $\lambda$ for two-step lasso). 
Right: $\ell_1$ error of the proposed method as $(\rho_1, \rho_2)$ vary (the upper half) and the symmetric Dirichlet method as $\rho = \rho_1=\rho_2$ varies (anti-diagonal elements in the lower half).}
\label{fig:simu}
\end{figure}

Figure~\ref{fig:simu} shows the sensitivity to hyperparameters for each method. The upper panel is for Scenario 1. From the plot, the proposed double spike Dirichlet prior yields uniformly better estimates than symmetric Dirichlet in terms of the $\ell_1$ error, which is expected since the double spike Dirichlet prior utilizes the structural information of the coefficients. Performances of two-step lasso largely depend on the choice of $\lambda$; in contrast, the proposed method is much less sensitive to the choice of $\rho_1$ and $\rho_2$. We also observe a monotonic pattern in the upper right heatmap of Figure~\ref{fig:simu}: the $\ell_1$ error tends to decrease with larger $\rho_1$ values and smaller $\rho_2$ values. Indeed, large $\rho_1$ and small $\rho_2$ values are desired in this scenario as the true $\beta^*$ does follow the sparse and partially constant structure. A limiting case is $\rho_1\to\infty$ and $\rho_2\to 0$, which amounts to $\beta_i=\gamma_i/|\gamma|$. Samples drawn from the double spike prior with such choices possess the exact structure, which is desirable when the exact structure holds. However, the numerical difference in $\ell_1$ error becomes indistinguishable for large enough $\rho_1$. 
The lower panel is for Scenario 2. The boxplot continues to show that the proposed method outperforms the others, although the sparse and partially constant assumption does not strictly hold. However, the monotone decreasing pattern in $\ell_1$ error as $\rho_1$ gets larger and $\rho_2$ gets smaller no longer holds (the lower right plot), which is not surprising as the true $\beta^*$ deviates from the exact structure. In both scenarios, the proposed method is not sensitive to $\rho_1$ and $\rho_2$ when $\rho_1$ exceeds certain thresholds. 

\subsection{Extended Comparisons} \label{sec:extended.sim}
We extend the previous section by varying the deviation level of $\beta^*$ from the strict sparse and partially constant structure. We also consider variants of methods in comparison. In particular, 
for the proposed method, instead of fixing $\theta = 1/K$, we put a prior $\text{Beta}(1, 0.5K)$ on $\theta$ and update it as well; for the two-step approach, instead of using lasso, one may use other adaptive shrinkage methods such as horseshoe~\citep{carvalho2010horseshoe}, leading to the method of two-step horseshoe.
We implement two-step horseshoe by using the default HS.var.select() function from the R package \textbf{horseshoe}, which selects a variable if its marginal credible interval does not contain 0.

We create 10 different $\beta^*$ values corresponding to different deviations from partial constancy. Starting with the $\beta^*$ vector in the second scenario, which uses $\beta_1=0.3089$ and distributes a total of 0.05 weight to the last $K-3$ elements, we vary $\beta_2$ from $0.2$ to $0.3$. The value of $\beta_3$ is determined by the simplex constraint. We choose this range for $\beta_2$ to ensure that their closest point on $\{\Theta(s,K):s\le K\}$ is $(1/3, 1/3, 1/3, 0,\ldots,0)$, which can be checked empirically. The $\ell_1$ distance between the generated $\beta^*$ and its projection in $\{\Theta(s,K):s\le K\}$ ranges from 0.1244 to 0.3155; this is a wide interval and covers the derivation in Scenario 2 in Section~\ref{sec:two.scenarios} (0.1677).

Table~\ref{table:deviation} shows the average $\ell_1$ error of $\beta$ by our methods and competing methods when the level of deviation varies. We observe a consistent pattern that the proposed methods outperform the symmetric Dirichlet and the two-step approaches. For our methods, the two versions with fixed $\theta$ and random $\theta$ do not show significant differences relative to standard errors, although the fixed $\theta$ approach has a slightly smaller $\ell_1$ error for all cases. For the two-step approaches, two-step horseshoe has smaller $\ell_1$ error than two-step lasso, but it is still uniformly worse than the proposed methods. This adds to evidence of efficiency gain by adopting the proposed unified approach relative to two-step alternatives. Besides, in our first real data application, we observed that two-step horseshoe failed to select any predictors.  
We next compare symmetric Dirichlet with the fixed $\theta$ version of our methods. Symmetric Dirichlet leads to uniformly larger $\ell_1$ errors, with a widening margin as the deviation turns smaller. It is reassuring that the proposed method continues to outperform alternatives when the deviation is as large as 0.3155. A close inspection of the symmetric Dirichlet method in one simulated data set indicates that it assigns in total more than 0.15 weight to the near-zero components of $\beta$, much larger than the true weight of 0.05, while our method assigns 0.007.  Therefore, our method appears to better maintain the sparsity than the symmetric Dirichlet. For a non-sparse $\beta^*$, which is not particularly interesting in the high-dimensional setting considered here, symmetric Dirichlet may become competitive. 

\begin{table}[tbp]
\centering
\begin{tabular}{cccccc}
\hline
\multirow{2}{*}{Deviation} & \multicolumn{2}{c}{Proposed} & \multirow{2}{*}{Symmetric Dir}& \multicolumn{2}{c}{Two-step}\\
\cline{2-3} \cline{5-6}
 & $\theta=1/K$ & $\theta\sim Beta$ &  & lasso & horseshoe \\
\hline
0.3155& 0.420 (0.010)& 0.432 (0.009)& 0.591 (0.013)& 0.464 (0.016)& 0.453 (0.017)\\
0.2943&	0.395 (0.011)& 0.414 (0.011)& 0.592 (0.015)& 0.448 (0.020)& 0.438 (0.018)\\
0.2731&	0.367 (0.012)& 0.387 (0.011)& 0.597 (0.016)& 0.436 (0.020)& 0.420 (0.019)\\
0.2518&	0.329 (0.012)& 0.355 (0.012)& 0.592 (0.016)& 0.409 (0.021)& 0.378 (0.020)\\
0.2306&	0.294 (0.012)& 0.318 (0.012)& 0.581 (0.019)& 0.384 (0.021)& 0.353 (0.021)\\
0.2094&	0.264 (0.012)& 0.292 (0.012)& 0.565 (0.020)& 0.354 (0.022)& 0.313 (0.018)\\
0.1881&	0.237 (0.012)& 0.261 (0.012)& 0.555 (0.021)& 0.328 (0.022)& 0.276 (0.017)\\
0.1669&	0.209 (0.012)& 0.225 (0.012)& 0.531 (0.020)& 0.308 (0.022)& 0.248 (0.017)\\
0.1457& 0.179 (0.010)& 0.195 (0.011)& 0.525 (0.021)& 0.295 (0.023)& 0.231 (0.018)\\
0.1244&	0.149 (0.008)& 0.166 (0.009)& 0.494 (0.019)& 0.277 (0.025)& 0.210 (0.018)\\
\hline
\end{tabular}
\caption{Comparison of the $\ell_1$ error of coefficient estimates for various levels of deviation from sparsity and partial constancy. Deviation of $\beta^*$ is defined as the $\ell_1$ distance between the generated $\beta^*$ and its projection in $\{\Theta(s,K):s\le K\}$.
Results are averaged over 100 simulations. Standard errors are reported in parentheses.}
\label{table:deviation}
\end{table}
In the appendix, we evaluate whether the signal-to-noise ratio would affect the comparisons among these methods. The results are consistent with what has been reported.

\section{Application to Survey Forecast Data}\label{Sec:realdata}
We use the European Central Bank's quarterly Survey of Professional Forecasters (SPF) data set to illustrate our method. The SPF collects information on the expected rates of inflation, real GDP growth, and unemployment in the euro area at several horizons, ranging from the current year to the longer term. Here we focus on making quarterly 1-year-ahead forecasts of euro-area real GDP growth (year-on-year percentage change) using the survey data from 1999Q1 to 2016Q2. This data set contains predictions from some forecasters of GDP growth rate for periods 1999Q3–2016Q4. The data preprocessing step, including a preliminary forecaster selection and missing data imputation procedure, follows \cite{diebold2019machine}. In total, we have 70 surveys, and each quarterly survey contains forecasts from $K=23$ forecasters. 

We calculate the out-of-sample root-mean-squared error to measure the performance of the proposed method, the Bayesian procedure using the noninformative symmetric Dirichlet prior, and two-step lasso. We also include the simple average method, which leads to robust and constantly favorable performances hinted by the forecast combination puzzle. For the proposed method, we take the linear combination of individual forecasts as the final prediction, where the weights are determined by the posterior mean of the coefficient. The evaluation period starts from 2000Q1, and the root-mean-squared error is calculated for the forecasts of 68 quarters.

Except for the simple average, all other methods require selection of hyperparameters. For our method, we fix $\rho_1=K^2$ and $\rho_2=1/K$; 
we fix $\theta=0.05$, which is roughly $1/K$. Similarly to the preceding section, we also implement a random-$\theta$ strategy by giving it the Beta$(1, 0.5K)$ prior. For two-step lasso, at each period, we search for the optimal $\lambda$ among the same grid of values used in Section~\ref{sec:simu}, and choose the one that yields the smallest root-mean-squared error historically. 

Starting from period two and rolling forward, we obtain the forecasts for $t=3,\ldots, 70$ based on which the root-mean-squared error is calculated. For our method, the fixed-$\theta$ strategy yields an out-of-sample root-mean-squared error 1.441. The random-$\theta$ counterpart performs similarly, leading to an error of 1.444; for the proposed method, we focus on the fixed-$\theta$ strategy in the following, unless stated otherwise. 
The simple average gives 1.502, and two-step lasso gives 1.542. A closer look at the squared prediction errors reveals three outliers for all methods, which occurred in 2008Q2, 2008Q3, and 2008Q4 for the GDP growth over 2007Q4-2008Q4, 2008Q1-2009Q1, and 2008Q2-2009Q2, respectively. This corresponds to the financial crisis in 2008. After removing the three outliers, the out-of-sample root-mean-squared error is 1.014 for our method, 1.032 for the simple average method, and 1.058 for two-step lasso. The two-step lasso method, which chooses weights adaptively, does not outperform the simple average method.
The comparison between the proposed method and two-step lasso confirms the advantage of a joint approach versus a two-step alternative, at least in this data application. 

The inferred weights in this data application are very sparse. Taking the prediction at 2011Q2 as an example (i.e., we fit the model using data up to 2011Q1), the posterior mean of $|\gamma| = \sum_{i=1}^K\gamma_i$ is 1.12, and the posterior mean of $\beta$ has four non-negligible components: 0.40, 0.26, 0.09, 0.09. This observation is consistent across quarters; for example, the posterior mean of $|\gamma|$ when averaged over quarters is 1.13. We have also observed that the proposed method can differentiate highly correlated predictors. For example, using data up to 2011Q1 again, the predictor with the most prominent $\beta$ (whose posterior mean is 0.40) turns out to be highly correlated with the 19 predictors with negligible coefficient estimates: the correlation ranges from 0.83 to 0.93. Therefore, in this data application we do not observe that the proposed method selects an arbitrary candidate from a set of highly correlated predictors. We additionally conduct a simple simulation study in Appendix C to assess the proposed method in the presence of high correlation among predictors, which shows supporting evidence for this interesting observation.

\begin{figure}[tbp]
    \centering
    \includegraphics[width=0.7\textwidth]{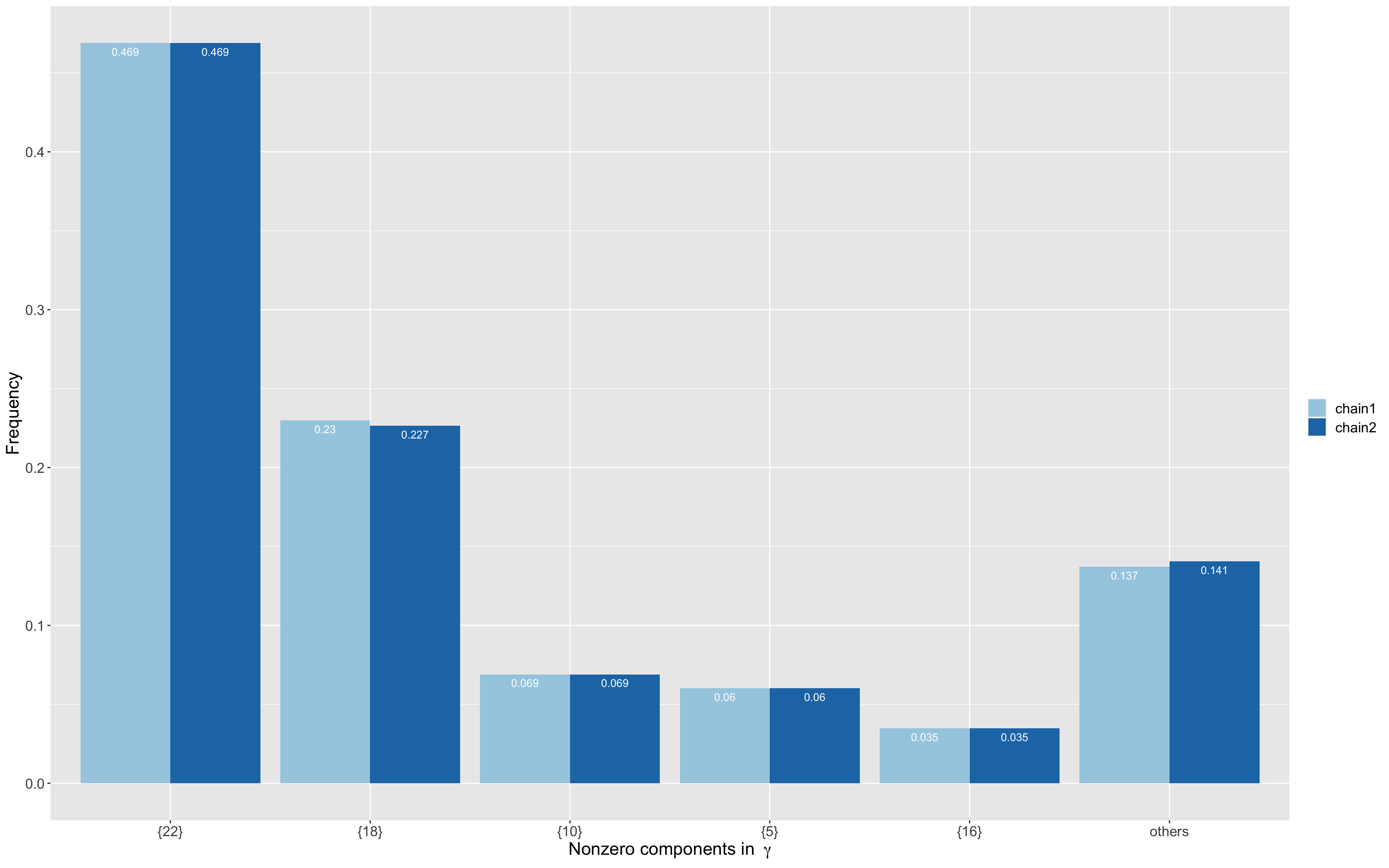}
    \caption{Frequencies of different $\gamma$ vectors in two parallel MCMC chains. The x-axis represents each $\gamma$ by its nonzero components; for example, ``\{22\}'' means only the 22th element of the corresponding $\gamma$ vector is nonzero. }
    \label{fig:spfmix}
\end{figure}

To assess the convergence for $\gamma$, we conduct Gelman and Rubin diagnostic~\citep{gelman1992inference} for each $\gamma_i$, $1\le i\le K$. This procedure runs two parallel Markov chains, each with a different initialization, and calculates a diagnostic statistic. 
For our method, the largest diagnostic statistic is $0.9999 < 1.1$, indicating that the MCMC is run long enough. Note that here we exclude the cases when the diagnostic statistic is reported as `NA', which occurs when $\gamma_i$'s empirical frequency after burn-in is close to the boundary of the parameter space (either 0 or 1). We further visualize the empirical frequency of different $\gamma$ vectors between the two parallel MCMC chains. Similarly to the logic behind the Gelman and Rubin diagnostic, if the sampler converges and mixes well, then the two parallel chains with different initializations should have similar frequencies for the same $\gamma$. In Figure~\ref{fig:spfmix}, we show the frequencies of the five most frequent $\gamma$ vectors and group all other $\gamma$'s into a combined category. The two frequency distributions for $\gamma$'s are close, again suggesting good mixing and convergence.

\section{Improving Random Forests Predictions}
\label{sec:RF}
In this section, we use the concrete data set from the UCI repository to demonstrate how the proposed method improves random forests. 
Random forests have been considered one of the most popular supervised learning methods. 
The high dimension of trees and the simple average structure in random forests (and also bagging) make the proposed method particularly well suited for further improvement. This will lead to structured random forests aided by double spike Dirichlet priors. The induced adaptive weighting effect that takes into account learners' performance has also been seen in other ensemble methods such as boosting~\citep{freund1997decision}, but unlike boosting, which considers learners in a sequential manner, our structured weighting jointly assesses all learners with additional shrinkage toward equal weights.

Concrete is a critical material in civil engineering. The concrete compressive strength is a function of age and ingredients. With the concrete data set \citep{yeh1998modeling}, we use eight input features, namely cement (kg/m$^3$), fly ash (kg/m$^3$), blast furnace slag (kg/m$^3$), water (kg/m$^3$), superplasticizer (kg/m$^3$), coarse aggregate (kg/m$^3$), fine aggregate (kg/m$^3$) and age of testing (days) to predict the compressive strength of concrete. The data set contains 1030 samples. We randomly sample 515 samples as the training set and use the rest for testing. We first construct random forests and obtain individual tree predictions using the \textbf{randomForest} package \citep{liaw2002classification} in R with the default settings for all arguments. This corresponds to $K=500$ trees. We then treat the 500 trees as 500 predictors and feed them into the proposed Bayesian procedure. Since
our simulations suggest the proposed method is not sensitive to hyperparameters, we set $\rho_1=K^2$, $\rho_2=1/K$, and $\theta=0.2$. Here we set $\theta$ larger than $1/K$ in order to leverage a larger number of trees for variance reduction, while still being more parsimonious than the original forests. 
We also implement the fully Bayesian version of our method by placing a Beta$(1,0.5K)$ prior on $\theta$. 
Since the number of predictors $K$ is large, we increase the number of iterations to 30000 (from the default value of 20000) and discard the first 20000 as burn-in. We then calculate the posterior mean weight for each tree and use this weighted sum as our proposed prediction. The out-of-sample root-mean-squared error is calculated on the test set. 

\begin{figure}[tbp]
\centering
\includegraphics[width =0.9\linewidth]{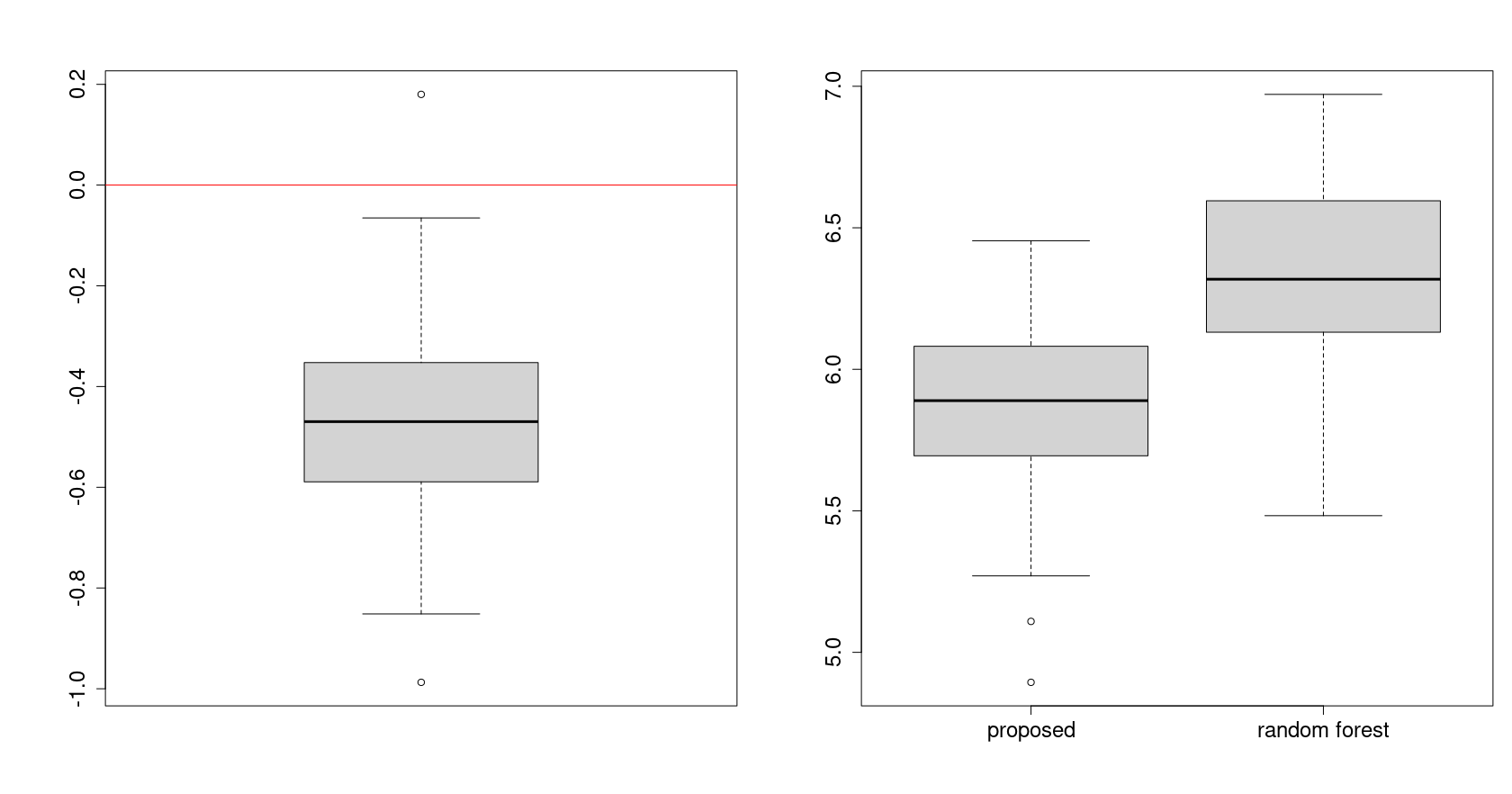}
\caption{Comparison of the out-of-sample root-mean-squared error of the proposed method and the random forests based on 100 replications. Left: difference of prediction error of the proposed method subtracting that of random forests. Right: prediction error of each method.}
\label{fig:RFoutrmse}
\end{figure}

Figure~\ref{fig:RFoutrmse} shows the out-of-sample root-mean-squared errors of the proposed method and the original random forests method, as well as their differences. The results are based on 100 replicates of the above procedure, each time generating a new train-test split. We can see that our method outperforms random forests in almost every train-test split, leading to a smaller out-of-sample root-mean-squared error. The average out-of-sample root-mean-squared error over 100 replications of our method is 5.870,
reducing the original random forests' 6.332 substantially by 7.30\%. For the full Bayesian version of our method, the improvement becomes 5.37\%, reducing the error from 6.332 to 5.992.

To gain insights on why the proposed method improves random forests, we extract one random forest consisting of 500 trees on one training-test set and calculate the number of trees that actually contribute to the final prediction, that is, the number of trees associated with $\gamma_i=1$, at each iteration after the burn-in period. It turns out that in this example, the number of contributing trees ranges from 10 to 18 out of 500 trees. In particular, only 13 trees are selected with a frequency higher than 0.2, only 29 trees are selected with a frequency higher than 0.05, and the vast majority of 393 trees are never selected. Thus, the proposed method leads to extreme parsimony when applied to random forests in this data set. 

\begin{figure}[tbp]
\minipage{0.5\textwidth}
  \includegraphics[width=\linewidth]{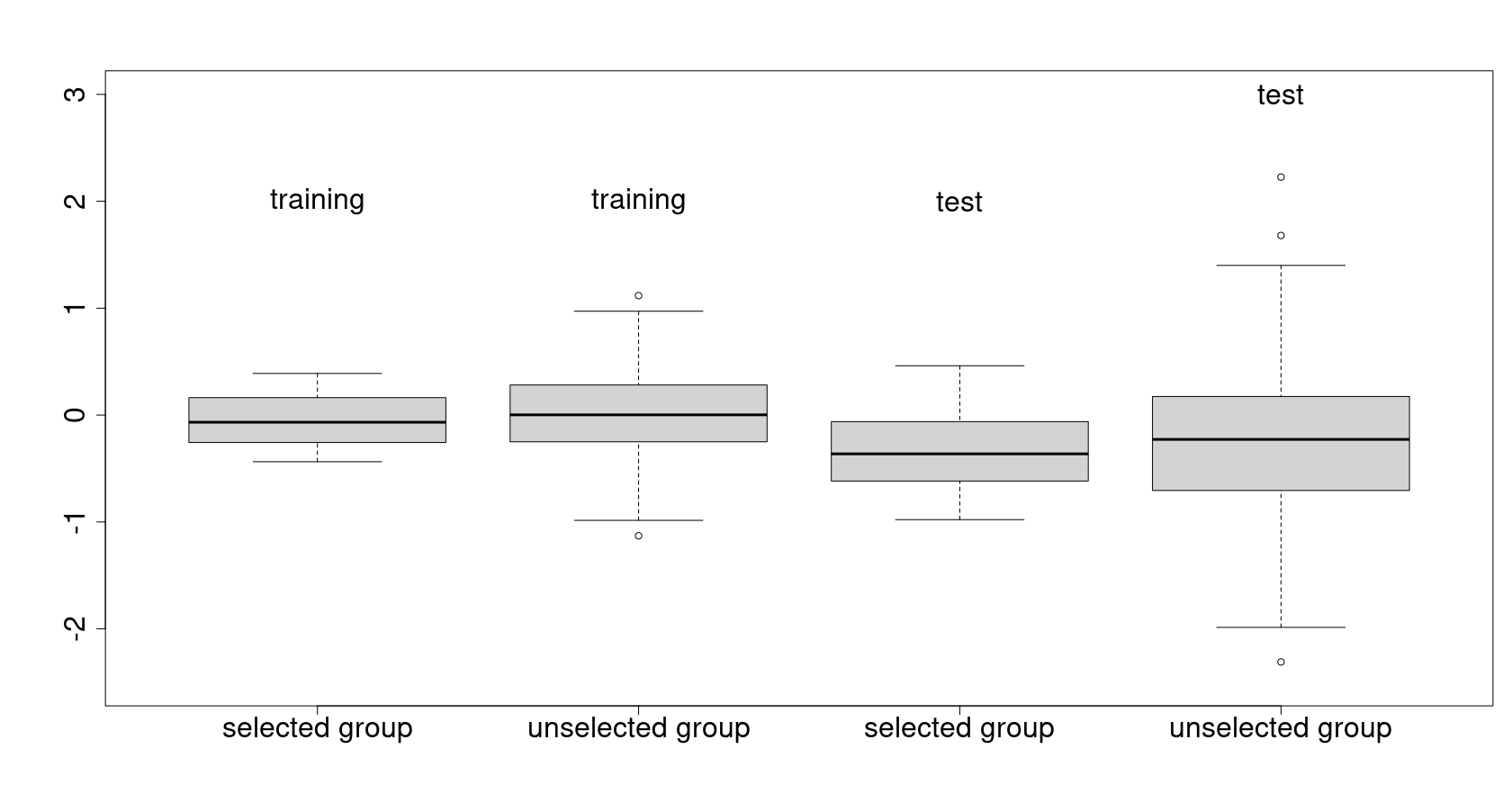}
\endminipage\hfill
\minipage{0.5\textwidth}
  \includegraphics[width=\linewidth]{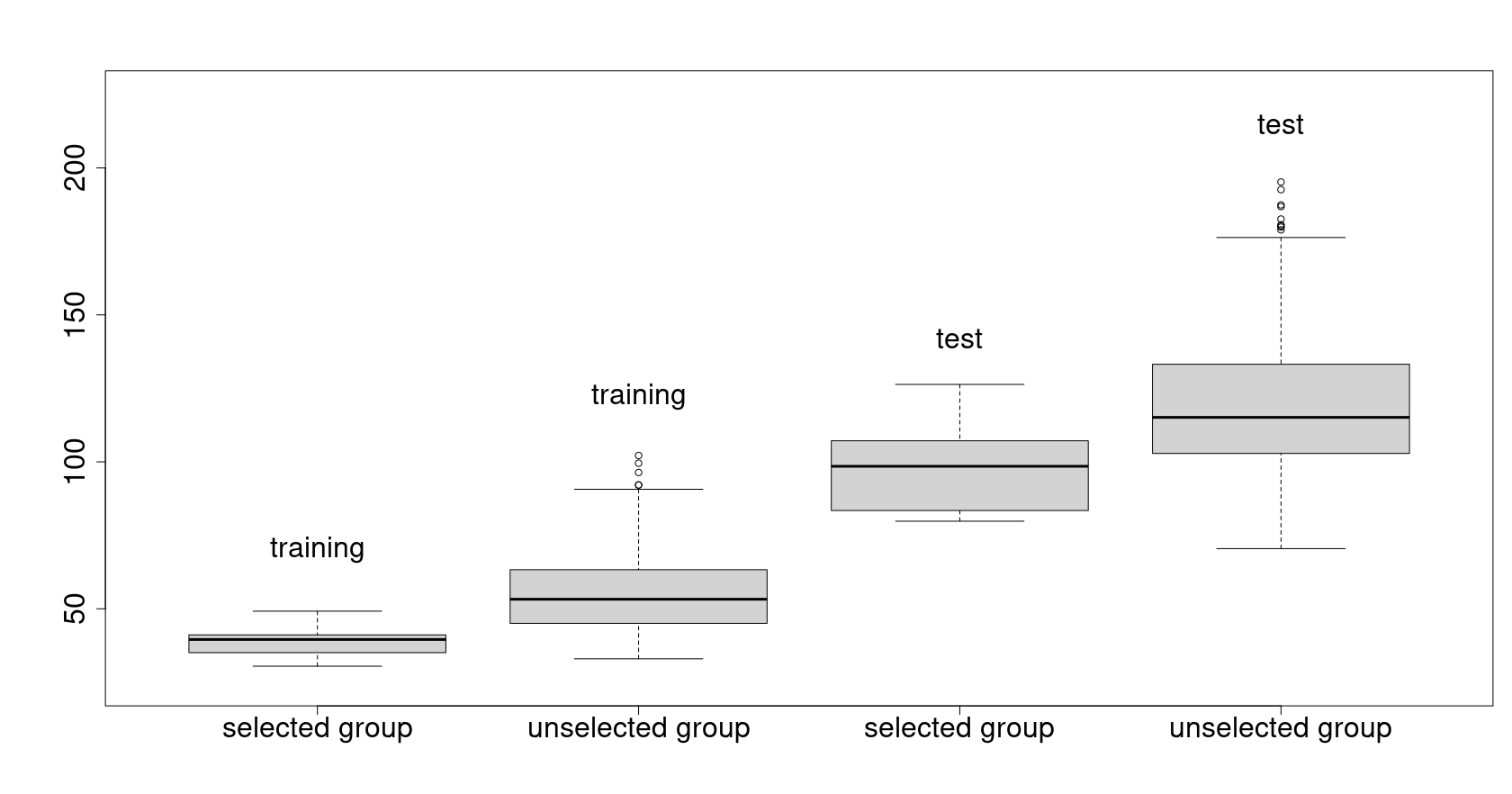}
\endminipage
\caption{Comparisons of the selected group and unselected group: biases (left) and variances (right).}
\label{fig:oneRF}
\end{figure}

We further term the 18 most frequently selected trees as the `selected group' and the rest as the `unselected group', and compare their performances to learn more insights. Taking the observed responses as the ground truth, for each tree, we calculate its averaged bias and variance over the samples. Figure~\ref{fig:oneRF} shows that the selected group generally has smaller variance than the unselected group, and larger bias appears more often in the unselected group than in the selected group. These observations hold for both the training and test sets. Therefore, the proposed method indeed selects a group of better predictors and generalizes well to the test set. Interestingly, the selected group does not necessarily consist of the best 18 individual trees, which are those trees yielding the  smallest mean-squared errors on the training set; in fact, the simple average of them gives a worse out-of-sample root-mean-squared error than the proposed method. Looking further into the selected group and the best 18 individual tree group suggests that this might be due to the larger correlation among members in the best 18 individual trees, causing them to bias toward the same direction and thus hampering the prediction. Indeed, we summarize tree correlations for each group by averaging all pairwise tree correlations in that group, where the pairwise tree correlation is defined as the correlation between the pair's prediction biases, and find that the selected group has an average pairwise correlation of 0.16, while the best 18 individual trees group has 0.23. This case study indicates that the proposed method accounts for the dependence across individual trees, which appears to be crucial to improve the prediction.

\section{Conclusions}
In this paper, we propose a novel class of double spike Dirichlet priors for structured high-dimensional probability simplexes. In particular, we accommodate the structure of sparsity and partial constancy, abstracting and extending a useful concept that emerged as the so-called `forecast combination puzzle' in economics to a broader class of applications. The proposed priors lead to a Bayesian procedure for estimating weight parameters. We design an efficient MCMC algorithm for posterior sampling, enabling inference and uncertainty quantification. We establish posterior concentration properties for the proposed priors under mild regularity conditions and proper choice of hyperparameters, which are adaptive to the unknown sparsity level. Simulations show that our method achieves a smaller estimation error and is more robust and flexible than its competitors in both ideal and less ideal scenarios. We resort to two applications to assess the performance of the proposed method using real world data. In the context of forecast combination, the proposed method exhibits superior predictive performance in an application to the European Central Bank Survey of Professional Forecasters data set; it further improves the prediction of random forecasts by more than 7\% using a UCI data set, suggesting its wide applicability building on the success and popularity of random forests. 

There are several interesting directions building on the proposed method. Firstly, our method can be extended to incorporate group structures. This can be achieved by placing a hierarchical prior on the binary vector $\gamma$ in~\eqref{eq:DSD}:
\begin{equation*}
\begin{split}
    \gamma_i&=\gamma_{g(i),1}\cdot \gamma_{i,2},\\
    \gamma_{g(i),1}& \overset{\text{independent}}{\sim} \text{Bernoulli}(\theta_G),\\
    \gamma_{i,2}&\overset{\text{independent}}{\sim} \text{Bernoulli}(\theta),
\end{split}
\end{equation*}
where $g_i$ is the group that $\beta_i$ falls into. Here $\theta_G$ represents one's prior belief on the group sparsity, and $\theta$ represents the prior belief on the within-group sparsity when group $g_i$ receives nonzero weights. In this formulation, $\gamma_i\neq0$ if and only if it receives nonzero weight ($\gamma_{i,2}\neq0$) within a nonzero group ($\gamma_{g(i),1}\neq0$). Such spike-and-slab priors on variable groups are commonly used to achieve group sparsity in other settings~\citep{bai2022spike}.

Secondly, the sampling of high-dimensional binary vectors using ADSS may have the potential pitfall of slow mixing~\citep{schafer2012monte}, particularly when the number of predictors is large. This is mitigated by the strong structure of the considered parameter space as discussed in preceding sections, and one can additionally increase the number of MCMC iterations for larger $K$ as in our example to improve random forests. In conventional sparse Bayesian variable selection, \cite{yang2016computational} showed under mild conditions certain random walk type of Metropolis-Hastings algorithm can mix rapidly. While our posterior sampling algorithm seems to work well in practice, it is interesting to theoretically investigate whether or not it enjoys such rapid mixing in the presence of structured simplexes.

Finally, non-negative weights on the simplex that our priors are supported on are widely adopted in many applications, such as forecast combination in economics and model selection. For example, the induced convex combination is shown to be more stable than its non-constrained counterpart~\citep{conflitti2015optimal} and remains a popular choice in model aggregation~\citep{bunea2007aggregation,polley2010super}. Nevertheless, an interesting future direction would be to relax the non-negative constraint of a simplex to mixed signs. 


\section*{Acknowledgements}
We thank David Dunson and Marina Vannucci for their helpful comments on an earlier version of this paper, and Minchul Shin for sharing the preprocessed SPF data used in Section~\ref{Sec:realdata}. We thank the action editor and three anonymous reviewers for constructive comments that helped to improve the paper. This work was partially supported by the grant DMS-2015569 and DMS/NIGMS-2153704 from the National Science Foundation, and a Ken Kennedy Institute 2020/21 Oil \& Gas High Performance Computing Conference Graduate Fellowship.



\appendix
\section*{Appendix A.}
This section contains the proofs of Proposition~\ref{lemma}, Proposition~\ref{lem2} and Theorem~\ref{maintheorem}. 

\begin{proof}(Proof of Proposition~\ref{lemma})
Let $A_i$ be drawn independently from $\text{Gamma}(\alpha,1)$ and $\pi_i=A_i/\sum_{i=1}^K A_i$, then it is well known that $(\pi_1,\ldots, \pi_K) \sim \text{Dir}(\alpha,\ldots, \alpha)$. Note that $\pi_{(K-1)}/\pi_{(K)}=A_{(K-1)}/A_{(K)}$, and the event $\{A_{(K-1)}\le t A_{(K)}\}$ for given $t\in (0,1)$ can be partitioned into $K$ disjoint equal-probability events as follows:
$$
\{A_{(K-1)}\le t A_{(K)}\}=\bigcup_{k=1}^K\{A_i\le t A_k, 1\le i \le K \text{ and } i \neq k\}.
$$
Therefore,
\begin{equation}\label{lemma:eq}
    \begin{split}
        \text{pr}(\pi_{(K-1)}\le t\pi_{(K)})&=\text{pr}(A_{(K-1)}\le t A_{(K)})\\
        &=K \text{pr}(A_1\le t A_K, A_2\le t A_K,\ldots, A_{K-1}\le t A_K)\\
        &=K E\left\{\text{pr}(A_1\le t A_K, A_2\le t A_K,\ldots, A_{K-1}\le t A_K|A_K)\right\}\\
        &=K E\left\{F^{K-1}(tA_K)\right\},
    \end{split}
\end{equation}
where the expectation in the last two steps is with respect to $A_K \sim \text{Gamma}(\alpha,1)$, and $F(\cdot)$ is the cumulative distribution function of $\text{Gamma}(\alpha,1)$. When $t\in (0,1)$, it follows that $F(tx)\ge t^{\alpha}F(x)$ for any $x>0$. This is because
\begin{equation*}
    \begin{split}
        F(tx)&=\int_0^{tx}\frac{1}{\Gamma(\alpha)}z^{\alpha-1}e^{-z}dz\\
             &\overset{y=z/t}{=}t^\alpha\int_{0}^{x}\frac{1}{\Gamma(\alpha)}y^{\alpha-1}e^{-ty}dy\geq t^\alpha\int_{0}^{x}\frac{1}{\Gamma(\alpha)}y^{\alpha-1}e^{-y}dy\\
             &=t^\alpha F(x).
    \end{split}
\end{equation*}
Combining this inequality with \eqref{lemma:eq}, we obtain that
$$
\text{pr}(\pi_{(K-1)}\le t\pi_{(K)})\ge Kt^{\alpha(K-1)}E\left\{F^{K-1}(A_K)\right\}.
$$
This completes the proof by noting that $E\left[F^{K-1}(A_K)\right]=\int_0^{\infty}F^{K-1}(x)dF(x)=1/K$.
\end{proof}

\begin{proof}(Proof of Proposition~\ref{lem2})
Under the proposed prior, a lower bound on the probability that $N\ge s$ is the probability that $\gamma$ has $s$ nonzero components and $\beta$ has $s$ components greater than $1/(2s)$. Due to exchangeability, this is simply $$\binom{K}{s} \theta^s(1-\theta)^{K-s}\cdot\text{pr}(\beta_i>1/(2s), \text{ for all }i\le s|\gamma_i=1 \text{ if }i\le s \text{ and }0 \text{ otherwise}).$$
Given that $\gamma_i=1 \text{ if }i\le s \text{ and }0 \text{ otherwise}$, for $i\le s$, $E(\beta_i)=\rho_1/(s\rho_1+(K-s)\rho_2)\in (1/(s+\rho_1^{-1}), 1/s)$ if $\rho_2\le 1/K$, and Var$(\beta_i)<(s^2\rho_1)^{-1}$. 
Under the condition $\rho_1\ge 5/s$, we have $1/(s+\rho_1^{-1})\ge 5/(6s)$. By union bound and Chebyshev inequality, \begin{equation*}
\begin{split}
    \text{pr}(\beta_i<1/(2s), \text{for some }i\le s)&\le s\cdot\text{pr}(\beta_i-E(\beta_i)<1/(2s)-1/(s+\rho_1^{-1}))\\
    &\le s\cdot \text{pr}(|\beta_i-E(\beta_i)|>1/(s+\rho_1^{-1})-1/(2s))\\
    &\le 9s/\rho_1.
\end{split}
\end{equation*}
Therefore,
\begin{equation*}
\begin{split}
\text{pr}(N\ge s)&\ge \binom{K}{s} \theta^s(1-\theta)^{K-s}\cdot\text{pr}(\beta_i\ge 1/(2s) \text{ for all }i\le s)\\
&\ge \binom{K}{s} \theta^s(1-\theta)^{K-s}(1-9s/\rho_1).
\end{split}
\end{equation*}
This completes the proof. 
\end{proof}

\begin{proof}(Proof of Theorem~\ref{maintheorem})
Denote by $p_\beta$ the density of $N(X\beta, I)$ distribution, and the corresponding likelihood ratio by
$$
\Lambda_{\beta,\beta^*}(Y)=\frac{p_\beta}{p_{\beta^*}}(Y)=\exp\left\{-\frac{1}{2}\left\Vert X(\beta-\beta^*)\right\Vert_2^2+(Y-X\beta^*)^\T X(\beta-\beta^*)\right\}.
$$
For any subset $B \subset \Theta$, by the Bayes formula, the posterior probability mass of its complement $B^c$ is
\begin{equation*}\label{post}
\begin{split}
 \Pi(B^c|Y)&=\frac{\int_{B^c}\Lambda_{\beta,\beta^*}(Y)g(\beta;\theta,\rho_1,\rho_2)d\beta}{\int_{\Theta}\Lambda_{\beta,\beta^*}(Y)g(\beta;\theta,\rho_1,\rho_2)d\beta}\\
 &=\frac{\int_{B^c}\Lambda_{\beta,\beta^*}(Y)g(\beta;\theta,\rho_1,\rho_2)d\beta}{\int_{B^c}\Lambda_{\beta,\beta^*}(Y)g(\beta;\theta,\rho_1,\rho_2)d\beta+\int_{B}\Lambda_{\beta,\beta^*}(Y)g(\beta;\theta,\rho_1,\rho_2)d\beta}\\
 &=:\frac{N_1}{N_1+N_2}.
\end{split}
\end{equation*}
Take $B=\{\beta \in \Theta:\left\Vert\beta-\beta^*\right\Vert_1\leq R\}$ for some $0<R<2$ that will be specified later. 
Without loss of generality, henceforth we assume that the first $s$ elements of $\beta^*$ are equal to $1/s$ and the rest are zero, while noting that the same argument applies for any $\beta^* \in \Theta(s,K)$.

We first consider to bound $N_2$. For any $0<r<R$, there holds
\begin{equation*}
    N_2\geq \int_{\left\Vert\beta-\beta^*\right\Vert_1<r}\Lambda_{\beta,\beta^*}(Y)D(\beta;\rho^*)\theta^s(1-\theta)^{K-s}d\beta,
\end{equation*}
where $D(\beta;\rho^*)$ is the Dirichlet distribution density with concentration parameter $\rho^*$ whose first $s$ elements are $\rho_1$ and the rest are $\rho_2$. Now under $\beta \sim \text{Dir}(\rho^*)$, for $1\leq i \leq s$, we have  $$E(\beta_i)=\frac{\rho_1}{\rho_1s+\rho_2(K-s)}=\frac{1}{s+(K-s)\rho_2/\rho_1}\in \left(1/s-s^{-2}K^{1-(\alpha_1+\alpha_2)},1/s\right).
$$
We discuss the case for $s=1$ and $s\ge 2$ separately. By Chebyshev's inequality, when $s\geq 2$,
we have for any $\v_1>0$,
$$
\text{pr}(|\beta_i-E(\beta_i)|>\v_1)\leq \frac{s^{-1}(1-s^{-1})}{sK^{\alpha_1}+(K-s)K^{-\alpha_2}+1}\frac{1}{\v_1^2}.
$$
In view of the union bound, with probability at least  $1-\left\{[sK^{\alpha_1}+(K-s)K^{-\alpha_2}+1]\v_1^2\right\}^{-1}$, we have $\sum_{i=1}^s|\beta_i-E(\beta_i)|\leq s\v_1$. Combining this with the triangle inequality yields
\begin{equation}\label{firstS}
    \sum_{i=1}^s|\beta_i-\beta_i^*|\leq \sum_{i=1}^s|\beta_i-E(\beta_i)|+ \sum_{i=1}^s|E(\beta_i)-\beta_i^*|\leq s\v_1+s^{-1}K^{1-(\alpha_1+\alpha_2)}.
\end{equation}
For $i>s$, under $\beta \sim \text{Dir}(\rho^*)$, the marginal distribution of $\sum_{i>s}\beta_i$ is $\text{Beta}\left(\rho_2(K-s), \rho_1s\right)$. Therefore, 
$$
E\left(\sum_{i>s}\beta_i\right)=\frac{\rho_2(K-s)}{\rho_2(K-s)+\rho_1s} \in \left(0, s^{-1}K^{1-(\alpha_1+\alpha_2)}\right).
$$
Again, by Chebyshev's inequality, for any $\v_2>0$,
\begin{equation*}
    \text{pr}\left(\left|\sum_{i>s}\beta_i-E(\sum_{i>s}\beta_i)\right|>\v_2\right) \leq  s^{-2}K^{1-(2\alpha_1+\alpha_2)}\v_2^{-2}.
\end{equation*}
Following a similar argument as used to derive \eqref{firstS}, we obtain that with probability at least $1-s^{-2}K^{1-(2\alpha_1+\alpha_2)}\v_2^{-2}$,
\begin{equation}\label{lastK-s}
    \sum_{i>s}|\beta_i-\beta_i^*|=\sum_{i>s}\beta_i
    \leq \left|\sum_{i>s}\beta_i-E\left(\sum_{i>s}\beta_i\right)\right|+E\left(\sum_{i>s}\beta_i\right)\leq \v_2+s^{-1}K^{1-(\alpha_1+\alpha_2)}.
\end{equation}
Combining \eqref{firstS} and \eqref{lastK-s}, with probability at least  $$1-[sK^{\alpha_1}+(K-s)K^{-\alpha_2}+1]^{-1}\v_1^{-2}-s^{-2}K^{1-(2\alpha_1+\alpha_2)}\v_2^{-2},$$ we have
\begin{equation}\label{L1error}
    \left\Vert\beta-\beta^*\right\Vert_1\leq s\v_1+\v_2+2s^{-1}K^{1-(\alpha_1+\alpha_2)}.
\end{equation}
Take $r=s\v_1+\v_2+2s^{-1}K^{1-(\alpha_1+\alpha_2)}<R$. Define the event $\tau_0=\{\left\Vert X^\T(Y-X\beta^*)\right\Vert_{\infty}\leq \delta\}$. Conditional on $\tau_0$, $N_2$ is lower bounded by a constant:  
\begin{equation}\label{LN2}
\begin{split}
    N_2\geq &\int_{\left\Vert\beta-\beta^*\right\Vert_1<r}\exp\left\{-\frac{1}{2}\left\Vert X(\beta-\beta^*)\right\Vert_2^2-\delta\left\Vert\beta-\beta^*\right\Vert_1\right\}D(\beta;\rho^*)\theta^s(1-\theta)^{K-s}d\beta\\
    \geq &\exp\left(-\frac{1}{2}\left\Vert X\right\Vert^2 r^2-\delta r\right)\theta^s(1-\theta)^{K-s}\\
    & \quad \cdot \left\{1-\left[sK^{\alpha_1}+(K-s)K^{-\alpha_2}+1\right]^{-1}\v_1^{-2}-s^{-2}K^{1-(2\alpha_1+\alpha_2)}\v_2^{-2}\right\}\\
     =: & L_{N_2}.
\end{split}
\end{equation}
Hence, conditional on $\tau_0$, it holds that
\begin{equation}\label{primLB}
    \Pi(B^c|Y)\leq \frac{N_1}{N_1+L_{N_2}}. 
\end{equation}


Let $\mathbbm{1}{\tau_0}$ and $\mathbbm{1}{\tau_0^c}$ indicate the occurrence of events $\tau_0$ and $\tau_0^c$, respectively. Note that
\begin{equation}\label{Epi}
    E_{\beta^*}\left\{\Pi(B^c|Y)\right\}=E_{\beta^*}\left\{\Pi(B^c|Y)\mathbbm{1}{\tau_0}\right\}+E_{\beta^*}\left\{\Pi(B^c|Y)\mathbbm{1}{\tau_0^c}\right\}\leq E_{\beta^*}\left\{\Pi(B^c|Y)\mathbbm{1}{\tau_0}\right\}+E_{\beta^*}(\mathbbm{1}{\tau_0^c}).
\end{equation}
It follows from \eqref{primLB} that 
\begin{equation}\label{EpiTau}
\begin{split}
   E_{\beta^*}\left\{\Pi(B^c|Y)\mathbbm{1}{\tau_0}\right\} &\leq E_{\beta^*}\left[\left(1-\frac{L_{N_2}}{N_1+L_{N_2}}\right)\mathbbm{1}{\tau_0}\right]\\
   &\leq 1-L_{N_2}\left\{E_{\beta^*}\left(\frac{1}{N_1+L_{N_2}}\right)-E_{\beta^*}\left(\frac{\mathbbm{1}{\tau_0^c}}{N_1+L_{N_2}}\right)\right\}\\
   &\leq 1-\frac{L_{N_2}}{E_{\beta^*}(N_1)+L_{N_2}}+E_{\beta^*}(\mathbbm{1}{\tau_0^c}),
\end{split}
\end{equation}
where the last line is a direct application of Jensen's inequality. 
Take $\delta=2\left\Vert X\right\Vert\surd{\log K}$. By Lemma 4 in \cite{castillo2015bayesian}, $E_{\beta^*}(\mathbbm{1}{\tau_0^c})\leq 2/K$. This combined with \eqref{Epi} and \eqref{EpiTau} yields
\begin{equation*}
    E_{\beta^*}\left\{\Pi(B^c|Y)\right\}\leq 1-\frac{L_{N_2}}{E_{\beta^*}(N_1)+L_{N_2}}+\frac{4}{K}.
\end{equation*}

It remains to show $E_{\beta^*}(N_1)=o(L_{N_2})$.
Take $\v_1=(\log K/sK^{\alpha_1})^{1/2}$, $\v_2=s\v_1$, and $R=s\log K/\{\Phi(s)\min(\left\Vert X\right\Vert, K^{\alpha_1/2})\}$. 
If $\alpha_1+\alpha_2\geq 1$,
\begin{equation*}
    sK^{\alpha_1}\v_1^2=\log(K),\quad s^{2}K^{-1+(2\alpha_1+\alpha_2)}\v_2^2=s^3K^{\alpha_1+\alpha_2-1}\log(K)>\log(K).
\end{equation*}
Therefore, $L_{N_2}$ defined in \eqref{LN2} satisfies that
\begin{equation*}\label{shortLN2}
L_{N_2}\geq \exp\left(-\frac{1}{2}\left\Vert X\right\Vert^2 r^2-\delta r\right)\theta^s(1-\theta)^{K-s}\frac{1}{2}
\end{equation*}
for sufficiently large $K$. 
On the other hand, by the definition of $\Phi(s)$, $E_{\beta^*}(N_1)$ is upper bounded by
$$
    E_{\beta^*}(N_1)=\int_B\exp \left(-\left\Vert X(\beta-\beta^*)\right\Vert_2^2\right)g(\beta;\rho_1,\rho_2,\theta)d\beta 
    \leq \exp\left(-\Phi^2(s)\left\Vert X\right\Vert^2R^2s^{-1}\right).
$$
Combining these two inequalities, a sufficient condition for $E_{\beta^*}(N_1)=o(L_{N_2})$ is
\begin{equation}\label{suffiCond}
\exp\left(-\Phi^2(s)\left\Vert X\right\Vert^2R^2s^{-1}\right) \ll 
\exp\left(-\frac{1}{2}\left\Vert X\right\Vert^2r^2-\delta r-\left\{s\log\left(\frac{1-\theta}{\theta}\right)-K\log(1-\theta)\right\}\right). 
\end{equation}

We next show that the left side of \eqref{suffiCond} is bounded by each component on the right side. With our specific choice of $\v_1$ and $R$, we obtain that
\begin{equation}\label{3inequal}
\begin{split}
    \frac{\left\Vert X\right\Vert^2(s\v_1)^2}{\Phi^2(s)\left\Vert X\right\Vert^2 R^2s^{-1}}&=\frac{\min(\left\Vert X\right\Vert^2, K^{\alpha_1})}{K^{\alpha_1}\log(K)}\leq \frac{1}{\log(K)}\\
    \frac{\delta s\v_1}{\Phi^2(s)\left\Vert X\right\Vert^2 R^2s^{-1}}&=\frac{2\min(\left\Vert X\right\Vert^2, K^{\alpha_1})}{\log(K)\left\Vert X\right\Vert\surd{K^{\alpha_1}}\surd{s}}\leq \frac{2}{\log(K)\surd{s}}\\
    \frac{s\log(K)}{\Phi^2(s)\left\Vert X\right\Vert^2 R^2s^{-1}}&=\frac{\min(\left\Vert X\right\Vert^2, K^{\alpha_1})}{\left\Vert X\right\Vert^2\log(K)}\leq \frac{1}{\log(K)}.
\end{split}
\end{equation}
When $\alpha_1/2+\alpha_2\geq 1$, it can be proved that $s^{-1}K^{1-(\alpha_1+\alpha_2)}\leq s\v_1$, and thus $r\leq 4s\v_1$ in view of \eqref{L1error}.  Choosing $\theta=t/K$ for $1\leq t\leq s$, as $K\to \infty$, it holds that $$
s\log\{(1-\theta)/\theta\}-K\log(1-\theta)\to s\log(K-t)-s\log(t)+t\leq s\log(K).
$$ Combining this with \eqref{suffiCond} and \eqref{3inequal}, we arrive at 
$E_{\beta^*}(N_1)=o(L_{N_2})$. 

The proof for the $s=1$ case is almost identical to the $s\ge 2$ case, with minor differences noted below. Under $\beta\sim\text{Dir}(\rho^*)$, $E(\beta_1)\in(1-K^{1-(\alpha_1+\alpha_2)},1)$. By Chebyshev's inequality, we have for any $\v_1>0$, 
$$
\text{pr}(|\beta_1-E(\beta_1)|>\v_1)\leq \frac{K^{1-(\alpha_1+\alpha_2)}}{K^{\alpha_1}+K^{1-\alpha_2}}\frac{1}{\v_1^2}.
$$
For $1<i<K$ under $\beta\sim\text{Dir}(\rho^*)$, the marginal distribution of $\sum_{i>1}\beta_i$ is Beta$(\rho_2(K-1), \rho_1)$. 
Using the same arguments for establishing \eqref{lastK-s} and \eqref{L1error}, we obtain that with probability at least $1-K^{1-(2\alpha_1+\alpha_2)}\v_2^{-2}$, 
$$
\sum_{i>1}|\beta_i-\beta_i^*|\le \v_2+K^{1-(\alpha_1+\alpha_2)},
$$
and that with probability at least $1-K^{1-(2\alpha_1+\alpha_2)}(\v_1^{-2}+\v_2^{-2})$,
$$
\lVert\beta-\beta^*\rVert_1\le \v_1+\v_2+2K^{1-(\alpha_1+\alpha_2)},
$$
for large enough $K$ and any $\v_2>0$. Take $r=\v_1+\v_2+2K^{1-(\alpha_1+\alpha_2)}<R$. We also have \eqref{primLB} but with a different $L_{N_2}=\exp(-\left\Vert X\right\Vert^2 r^2/2-\delta r)\theta(1-\theta)^{K-1}\left\{1-K^{1-(2\alpha_1+\alpha_2)}(\v_1^{-2}+\v_2^{-2})\right\}$. As \eqref{Epi}-\eqref{EpiTau} remain unchanged, to prove the theorem, we only need to show $E_{\beta^*}(N_1)=o(L_{N_2})$. 

Take $\v_1=\v_2=(2\log K/K^{\alpha_1})^{1/2}$ and $R=\log K/\{\Phi(1)\min(\left\Vert X\right\Vert, K^{\alpha_1/2})\}$. Since $\alpha_1+\alpha_2>1$, there holds $K^{1-(2\alpha_1+\alpha_2)}(\v_1^{-2}+\v_1^{-2})\le 1/\log K$. Therefore, for large enough $K$, the new $L_{N_2}$ is also lower bounded by
$$
L_{N_2}\ge \exp\left(-\frac{1}{2}\left\Vert X\right\Vert^2 r^2-\delta r\right)\theta(1-\theta)^{K-1}\frac{1}{2}.
$$
Noting that $r\le 3\v_1$ and choosing $\theta=1/K$, the rest of the proof remains exactly the same.
This completes the proof.
\end{proof}

\section*{Appendix B. Additional Simulations with Varying Signal-to-noise Ratio}
The signal-to-noise ratio (SNR) in Scenario 1 in Section~\ref{sec:two.scenarios} is 1.3. 
Here we conduct experiments with SNR varying from 0.6 to 2. This is done by adjusting the noise variance while fixing the design matrix and $\beta$. Similarly to Section~\ref{sec:two.scenarios}, we use default hyperparameter values for the proposed method, and give an advantage to the symmetric Dirichlet approach by reporting its best average $\ell_1$ error from a grid search for its hyperparameter. The two-step approaches, namely two-step lasso and two-step horseshoe, are also evaluated. Table~\ref{table:snr} shows that all methods benefit from a stronger signal, and the proposed method leads to the smallest $\ell_1$ error among all methods for all SNRs considered here.
\begin{table}[tbp]
    \centering
    \begin{tabular}{ccccc}
    \hline
     SNR & Proposed & Symmetric Dirichlet & Two-step lasso & Two-step horseshoe \\
    \hline
0.6&	0.366	(0.0261)&	0.879	(0.0263)& 0.443(0.0421)& 0.500 (0.0458)\\
0.8&	0.181	(0.0215)&	0.720	(0.0253)& 0.303(0.0334)& 0.275 (0.0380)\\
1.0&	0.090	(0.0154)&	0.581	(0.0228)& 0.201(0.0316)& 0.157 (0.0263)\\
1.2&	0.047	(0.0098)&	0.483	(0.0195)& 0.156(0.0291)& 0.105 (0.0221)\\
1.4&	0.029	(0.0058)&	0.414	(0.0159)& 0.129(0.0253)& 0.077 (0.0184)\\
1.6&	0.019	(0.0015)&	0.370	(0.0126)& 0.089(0.0222)& 0.075 (0.0179)\\
1.8&	0.018	(0.0006)&	0.334	(0.0091)& 0.069(0.0193)& 0.070 (0.0174)\\
2.0&	0.017	(0.0001)&	0.314	(0.0088) & 0.053(0.0173) & 0.070 (0.0174)\\
\hline
    \end{tabular}
    \caption{Comparison of the mean $\ell_1$ error of coefficient estimates of the proposed method, symmetric Dirichlet, and two-step methods at different signal-to-noise ratios. Results are based on 100 simulations. Standard errors are reported in parentheses.}
    \label{table:snr}
\end{table}

\section*{Appendix C. Additional Simulations with Correlated Predictors}
\begin{figure}[tbp]
\centering
\includegraphics[width=0.7\linewidth]{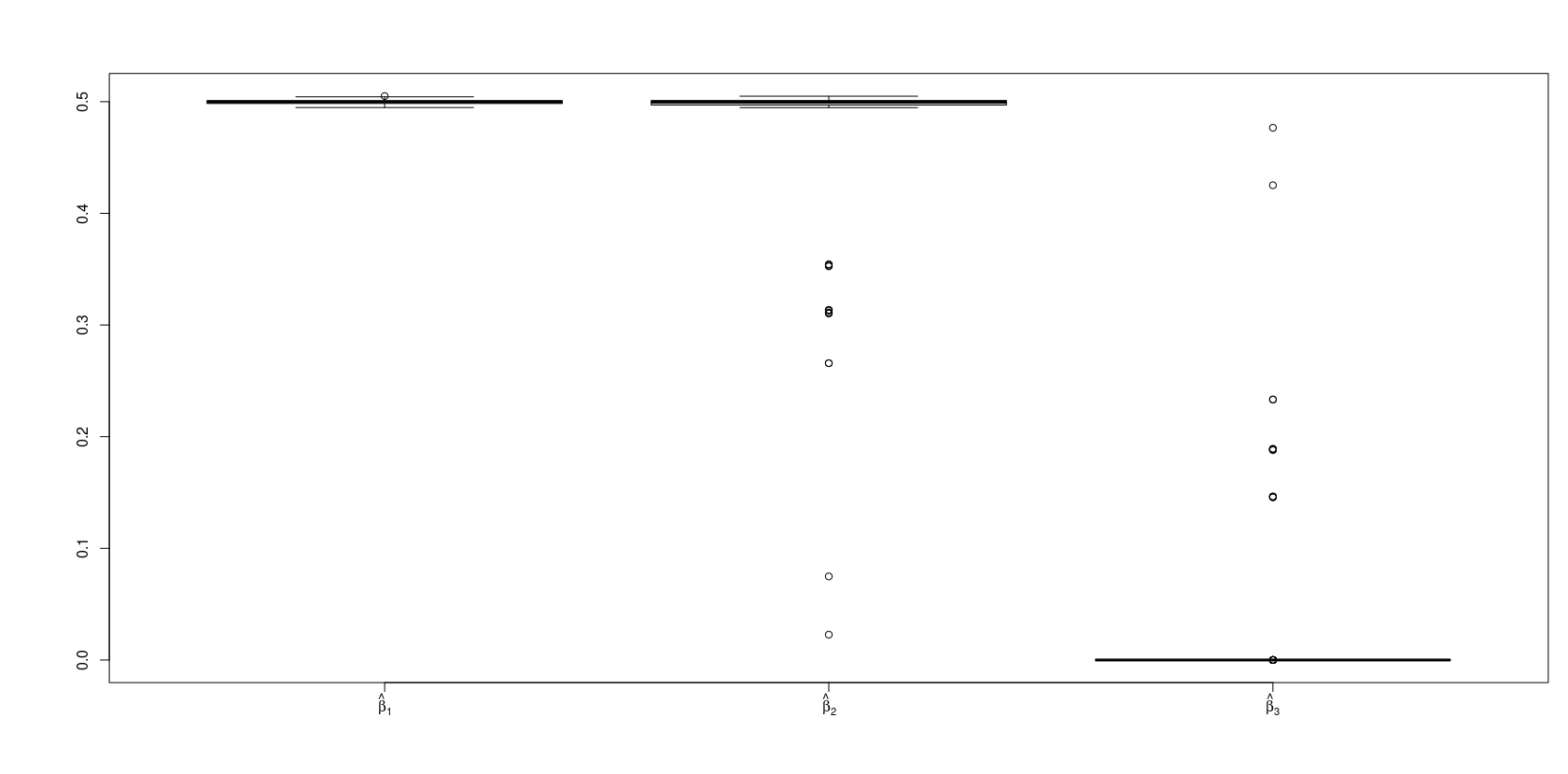}
\caption{Boxplot of posterior mean $\hat{\beta}_1$, $\hat{\beta}_2$, $\hat{\beta}_3$ in 100 replications.}
\label{fig:corrX}
\end{figure}

We set $n=80$, $K=40$, $s=2$, and assign nonzero weights to only the first two columns $\beta_1=\beta_2=0.5$. All columns of $X$ are independently drawn from $N(0, 3^2)$, with the exception that the second and the third columns are highly correlated with correlation 0.9. We apply the proposed method and calculate the posterior mean of $\beta$. This procedure is replicated for 100 times. Figure~\ref{fig:corrX} shows the boxplot of the posterior mean $\hat{\beta_i}$ for $i=1,2,3$. We can see that $\hat{\beta}_1$ concentrates tightly around 0.5, indicating good signal recovery for the first column. For the two correlated predictors, $\hat{\beta}_2$ concentrates around 0.5 and $\hat{\beta}_3$ around 0, both with a few outliers that come from the same replications. Therefore, our method is able to differentiate the two predictors most of the time, and is not arbitrarily selecting one out of the two highly correlated variables in this experiment. This is not surprising as under the exact structure the considered parameter space leads to a significant gap $1/s$ between the weights for true signals and those for null signals due to partial constancy, which eases the identification of true signals even under high correlation.

\bibliographystyle{apalike}
\bibliography{main}
\end{document}